# Observability of negative capacitance of a ferroelectric film: Theoretical predictions.

Eugene. A. Eliseev[1], Mykola E. Yelisieiev[2], Sergei V. Kalinin[3*] and Anna N. Morozovska[4†]

[1] Institute for Problems of Materials Science, National Academy of Sciences of Ukraine, Krjijanovskogo 3, 03142 Kyiv, Ukraine

[2] Taras Shevchenko National University of Kyiv, Volodymyrska street 64, Kyiv, 01601, Ukraine

[3] Department of Materials Science and Engineering, University of Tennessee, Knoxville, TN 37920

[4] Institute of Physics, National Academy of Sciences of Ukraine, 46, pr. Nauky, 03028 Kyiv, Ukraine

**Abstract**

We theoretically explore mechanisms that can potentially give rise to the steady-state and transient negative capacitance in a uniaxial ferroelectric film stabilized by a dielectric layer. The analytical expressions for the steady-state capacitance of a single-domain and poly-domain states are derived within Landau-Ginzburg-Devonshire approach and used to study the state stability vs. the domain splitting as a function of dielectric layer thickness. Analytical expressions for the critical thickness of the dielectric layer, polarization amplitude, equilibrium domain period and susceptibility are obtained and corroborated by finite element modelling. We further explore the possible effects of nonlinear screening by two types of screening charges at the ferroelectric-dielectric interface and show that if at least one of the screening charges is very slow, the total polarization dynamics can exhibit complex time- and voltage dependent behaviors that can be interpreted as an observable negative capacitance. In this setting, the transient negative capacitance effect is accompanied by almost zero dielectric susceptibility in a wide voltage range and low frequencies. These results may help to elucidate the observation of the transient negative capacitance in thin ferroelectric films, and identify materials systems that can give rise to the behavior.

## I. Introduction

---

[*] corresponding author, e-mail: sergei2@utk.edu
[†] corresponding author, e-mail: anna.n.morozovska@gmail.com

The concept of negative capacitance (**NC**) in ferroelectric nanostructures has captivated attention of the condensed matter physics community since early 70's and especially during last decades [1]. Beyond purely fundamental interest [2], it was proposed that NC can be crucial for applications in nanoelectronics [3, 4, 5] and information technology applications, since the negative differential capacitance can minimize drastically the Joule heat generated during the device operation [6, 7, 8, 9].

It is important to distinguish between a steady-state and transient NC in ferroelectric materials [6]. The origin of the steady-state intrinsic NC can be the unstable hysteresis branch of the ferroelectric polarization dependence on electric field appearing due to the double-well free energy potential of the polarization; and the transient NC can be related with some change in the electrostatic energy, e.g., during the domain wall motion in external field [10, 11], or with other changes in polarization dynamics [1]. The transient NC also can occur due to the delay between the polarization reversal process and its retarded compensation by screening charges [12].

A while ago, Salahuddin and Datta [13] proposed a simple model of a steady-state NC based on the phenomenological Landau approach and electrostatic consideration for the system "a single-domain ferroelectric film - dielectric layer", placed between ideally-conducting electrodes. The central premise of this model was to stabilize the unstable hysteresis branch of the ferroelectric polarization $P$ dependence on electric field $E$ near small values of polarization ($P = 0$). For this unstable branch, the dielectric susceptibility, defined as $dP/dE$, becomes negative. At the same time, from the thermodynamic viewpoint the measured dielectric permittivity, defined as $dP/dE_{ext}$, where $E_{ext}$ is an applied external field, should be positive in a thermodynamically stable state [14]. In order to overcome the instability of the paraelectric phase ($P = 0$) below Curie temperature they proposed to add a dielectric layer without the spontaneous polarization that destabilizes the ferroelectric phase ($P \neq 0$) due to internal depolarization field caused by the discontinuity of spontaneous polarization at the interface [see **Fig. 1(a)**].

In other words, Salahudin et al. [13] using the above simple model assumed that the "NC regions are ordinarily unstable and not observed in experiments. But placing such a capacitor in series with a positive capacitor stabilizes it by making the effective capacitance of the composite capacitor positive, provided its thickness is less than the critical thickness" of the ferroelectric film. The simple explanation (given in 2008) was rather an independent guess, while a fundamental ground had been given a while ago, in 2001. Actually, the polarization states and dielectric response of a ferroelectric covered by a dielectric layer were studied much earlier by Bratkovsky and Levanyuk [15], and their results play a significant role in understanding the physical phenomena in this system. In particular, it was shown that the dielectric response of the ferroelectric subsystem cannot be separated from the response of a whole system, because the dielectric susceptibility of the ferroelectric layer, determined as polarization derivative with respect to the internal electric field, $dP/dE$, is the nonlocal quantity related to the whole system and it inevitably depends on the properties of the dielectric layer. This nonlocal behavior is due to a long-range Coulomb interaction [15]. Note, that the internal field can be changed experimentally via a change of the external electric field, and the change of the total capacitance is an observable quantity.

Hoffmann et al. [16] noted that a steady-state NC could not be stabilized in a homogeneously polarized system, since the latter will be unstable towards domain formation due to the depolarization field energy. Bratkovsky and Levanyuk [15] consider the domain-like state of a ferroelectric film and derive the equation with respect to the external field in this state. Their consideration uses the so-called "Kittel model" of domain structure, where the ferroelectric domains with opposite polarization direction are separated by the infinitely thin straight domain walls. Using the Landau approach and the Kittel model as proposed by Kopal et al. [17], Zubko et al. [18] considered a ferroelectric/dielectric superlattice, aiming to establish a theory of the NC in a wide temperature range including multidomain states. Note, that the Kittel model gives a qualitative picture of the domain structure behavior in a ferroelectric film [19], but suffers from the following artifact. If one calculates the distribution of the internal electric field, it turns out that there are regions where the field significantly exceeds the coercive one (see e.g., pages 143 - 144 in Ref.[19]).

In order to avoid the artifact, the phenomenological Landau-Ginzburg-Devonshire (**LGD**) approach can be used. For instance, using the linearized LGD approach for uniaxial ferroelectrics, Chensky and Tarasenko [20] calculated the threshold of domain formation with diffuse domain walls. The nonlinear LGD approach allows to consider the mechanisms of domain wall broadening near the film surface in a self-consistent manner (see e.g., [21, 22]). Using the LGD approach and finite elements modelling (**FEM**) Park et al. [23], as well as Saha and Gupta [24] analyzed the in-field changes of the domain structure in a uniaxial ferroelectric heterostructure and demonstrated that the polarization derivative with respect to local electric field (or local potential), which includes the contribution from the internal depolarization field, can be negative. Using LGD approach for multiaxial ferroelectrics, Ponomareva et al. [25] reveal that the appearance of flux-closure domains and vortex-like domain structures can lead to the anomalous behavior (negative value) of the static dielectric susceptibility in ferroelectric nanostructures.

To parallel theoretical studies, a number of experimental results have been interpreted as the evidence of NC (see review [6] and Refs. therein). For instance, Khan et al [26] considered a bilayer consisting of a ferroelectric $Pb(Zr_{0.2}Ti_{0.8})O_3$ and a dielectric $SrTiO_3$ layers, and compared its dielectric permittivity with those of SrTiO$_3$ layer alone. Their argument for NC is based on the straightforward application of the serial capacitance equation to the experimental data. However, it should be noted that the series equation for the capacitance does not account for the spatial dispersion of polarization and corresponding dielectric permittivity. Note, that according to Bratkovsky and Levanyuk [15] the total permittivity, which is a non-local value, characterizes the whole system, but not its separate parts.

Later, treating the same system, Khan et al [27] have shown that a dynamic response of the bilayer demonstrates signature of transient negative dielectric susceptibility. Similar results were obtained by Hoffman et al [28] and Kim et al [29] for the capacitor containing an ultrathin HfO$_2$ layer. Note that Khan et al [27] have shown that the internal electric field can be separated experimentally.

Yadav et al. [30] considered epitaxial PbTiO$_3$/SrTiO$_3$ superlattices and reported about negative local susceptibility at the cores of vortex-like structures, directly observed by scanning transmission electron microscopy. Neumayer et al. [31] proposed a Van der Waals layered ferroelectric CuInP$_2$S$_6$ to achieve bulk NC through the high mobility of Cu ions between different wells in the multi-well potential of atomic structure. They claimed that observed hysteresis loops of local piezoelectric response with inverse slope can be interpreted as the evidence of NC.

From the above overview, the common theoretical model used to explore the emergence of the steady-state NC is that of a material with a double-well free energy potential with two stable and one unstable extreme points. The unstable point is characterized by a negative susceptibility, and may be stabilized by adding the parabolic energy landscape of a linear dielectric to the double-well potential [see **Fig. 1(a-c)**]. However, the key aspect of such models is the contribution of the interaction energy to the dielectric susceptibility of a ferroelectric layer with respect to local (internal) electric field, $dP/dE_F$. The treatment of this non-local term, its manifestation in local properties and macroscopic device-level responses varies between different models [see e.g., **Fig. 1(d-f)**]. The negative derivative $dP/dE_F$ is treated as evidence of the static NC. Let us underline that the NC can be achieved only in one element ("film") of the electric circuit, but not in the whole circuit ("film +layer").

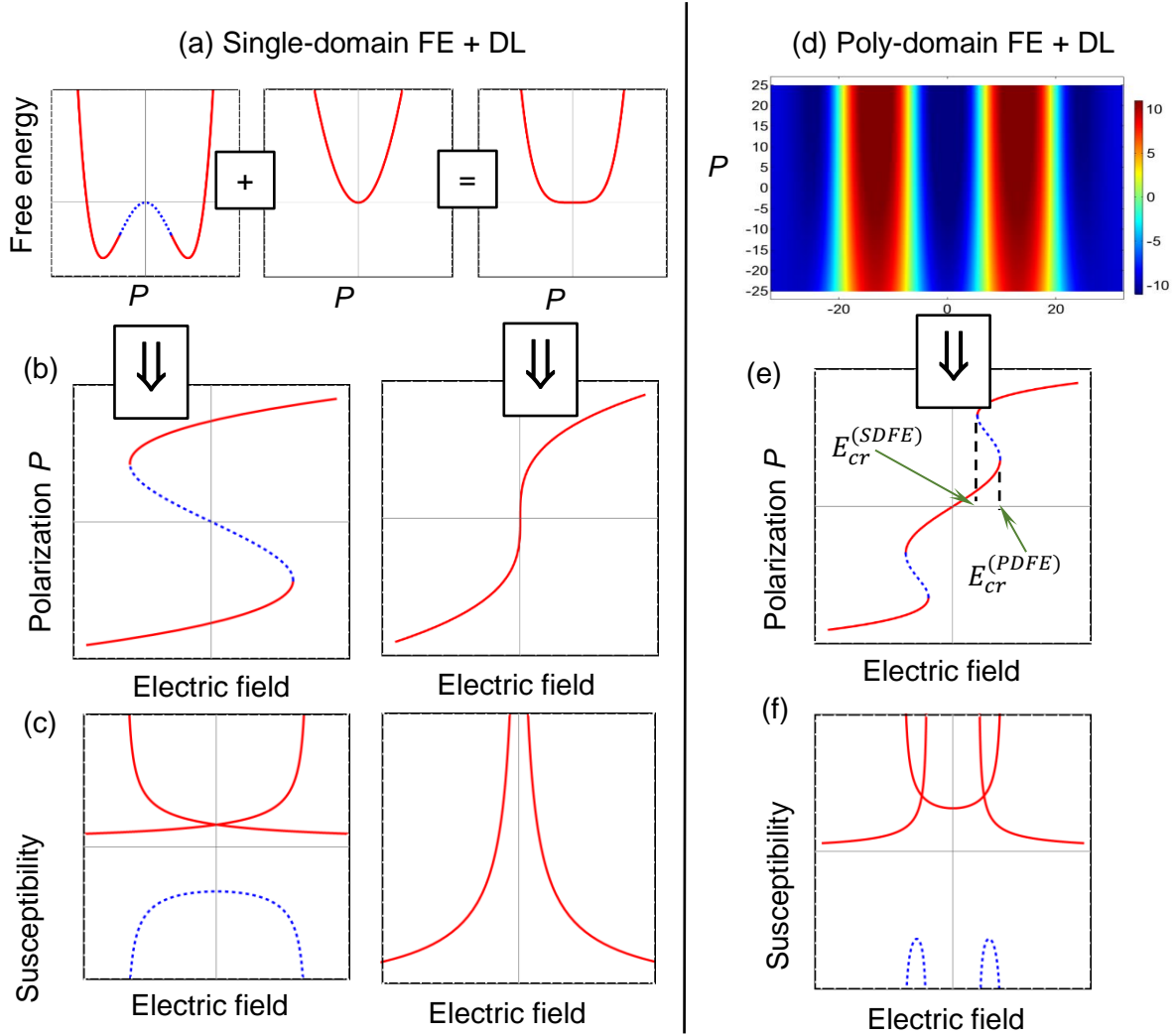

**Figure 1**. **(a)** A single domain state: free energy dependence on polarization for bulk ferroelectric (left), dielectric layer (center) and the combined single-domain ferroelectric (SDFE) – dead layer (DL) system (right). Polarization **(b)** and susceptibility **(c)** hysteresis curves for a bulk FE (left column) and for a FE film with a DL (right column). **(d)** Typical polarization distribution for a poly-domain ferroelectric (PDFE), corresponding polarization **(e)** and susceptibility **(f)** hysteresis curves. Solid red curves correspond to stable states, while dashed blues ones show the properties of unstable states.

Here, we aim to reexplore the fundamental models [presented in **Figs. 1**] and analyze the observability of the steady-state and transient NC, addressing both local and device level responses and static and dynamic aspects of this behavior. Based on this framework, we propose ***an alternative approach*** to the possible origin of the quasi steady-state transient NC from the slow ionic dynamic (or other subsystem coupled to fast ferroelectric polarization dynamics). Similar to previous treatments [13, 15], we first consider a single-domain ferroelectric layer covered by a dielectric layer and sandwiched between the electrodes. We derive the full GLD model of the systems explicitly including the interaction energy and obtain exact analytical expressions for the system capacitance. Note that this device-level response is not a sum of the inverse capacitance of the layers (**Section II**) corroborating results [15]. Using self-consistent LGD approach, we further consider the capacitance of the film with a domain structure, both analytically and using FEM, and corroborate results [15, 17, 18] (**Section III**). Finally, we explore the effect of a screening layer at the film surface, which consists of sluggish positive and negative free charges with different relaxation times and strongly nonlinear dependence of the charge densities on electric potential (**Section IV**). We conclude that the nonlinear ultra-slow dynamics of the screening charge can give rise to the quasi steady-state transient NC (**Section V**). Calculation details are given in **Supplemental Materials** [32].

## II. Capacitance in a Single-Domain State

As the baseline case, we consider a system consisting of the ferroelectric film covered by a dielectric layer placed between conducting electrodes [see **Fig. 2(a)**]. Note, that there are different origins of dielectric layers, also sometimes called "dead" or "passive" layers [33, 34, 35]. However, in most cases they are assumed to be ultrathin diffuse subsurface layers less than a few nanometers thick, where a spontaneous polarization is absent (or negligibly small) due to surface reconstruction, contamination, zero extrapolation length, and / or strong depolarization field. Irrespectively of the origin, a dead layer is typically approximated as a linear dielectric with a small thickness. The relative dielectric constant of the dielectric layer, $\varepsilon_e$, is usually quite high: $\varepsilon_e \sim$ (30 - 300).

The constant voltage $V$ is applied to the bottom electrode. For the sake of simplicity, here we consider a uniaxial ferroelectric with an out-of-plane spontaneous polarization $P_3$ directed along z-axis.

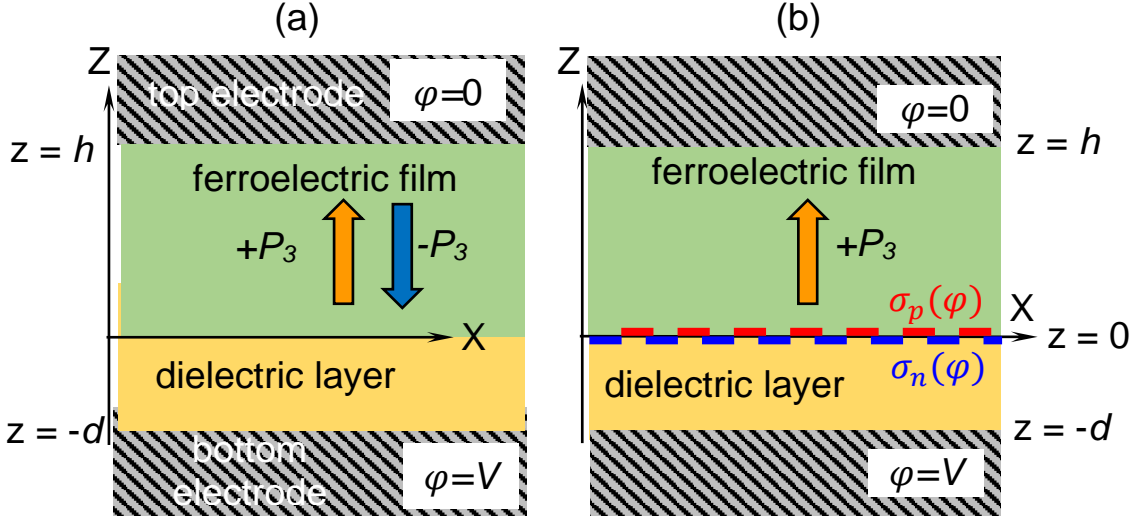

**Figure 2**. The schematics of the considered systems. **(a)** A uniaxial ferroelectric film of thickness $h$ and polarization vector $\boldsymbol{P}$ dielectric layer of thickness $d$, respectively, placed between the electrodes. The voltage $V$ is applied to the bottom electrode. **(b)** A top electrode, a ferroelectric film, a layer of positive and negative surface charges with densities, $\sigma_p(\phi)$ and $\sigma_n(\phi)$, a dielectric gap of width $d$, and a bottom electrode, which in principle, can provide a direct charge exchange with an ambient media.

The free energy of the system is assumed to follow the LGD form:

$$G = \int_{z\in[0,h]} \left(\frac{\alpha}{2}P_3^2 + \frac{\beta}{4}P_3^4 + \frac{\gamma}{6}P_3^6 + \frac{g_{11}}{2}\left(\frac{\partial P_3}{\partial z}\right)^2 + \frac{g_{44}}{2}\left[\left(\frac{\partial P_3}{\partial y}\right)^2 + \left(\frac{\partial P_3}{\partial x}\right)^2\right] - P_m E_m^{(i)} - \frac{\varepsilon_0 \varepsilon_b}{2} E_m^{(i)} E_m^{(i)}\right) d^3r - \int_{z\in[-d,0]} d^3r \frac{\varepsilon_0 \varepsilon_e}{2} E_m^{(e)} E_m^{(e)} \quad (1)$$

Here the first integral in Eq.(1) is the energy of a uniaxial ferroelectric film that includes the Landau expansion, polarization gradient energy and electrostatic contributions related to the internal electric field $\vec{E}^{(i)}$. The gradient coefficients $g_{11}$ and $g_{44}$ determine the correlation effects [36]. The coefficient $\alpha$ linearly depends on the temperature $T$, namely $\alpha = \alpha_T(T - T_c)$, where $T_c$ is a Curie temperature of a bulk uniaxial ferroelectric. The parameter $\varepsilon_b$ is a relative background permittivity of the ferroelectric [34, 37]. The second integral in Eq.(1) is the electrostatic energy of the dielectric layer that is defined by the electric field $\vec{E}^{(e)}$ inside the layer. Note that the polarization gradient energy and effects related with the internal electric field (including the background contribution) are key terms for producing further results of this work.

Mention that possible influence of elastic strains can be taken into account by renormalizing the coefficients $\alpha$ and $\beta$ in Eq.(1), as it was illustrated earlier for thin ferroelectric films [38]. This approach is enough rigorous in a single-domain state of the ferroelectric film [39], and corresponding analytical expressions, $\alpha^* = \alpha - \frac{2Q_{12}u_m}{s_{11}+s_{12}}$ and $\beta^* = \beta + \frac{Q_{12}^2}{s_{11}+s_{12}}$, where $Q_{ij}$ are electrostriction coefficients, $s_{ij}$ are elastic compliances and $u_m$ is the mismatch strain, are well-known (see e.g., Eq.(4) in [38]). However, the validity of the renormalization is questionable for a poly-domain configuration for which only numerical simulations are available [40]. The renormalization of LGD coefficients by elastic strains can be important for the quantitative description of the film polar properties and we will account for the effect in the FEM, which is sensitive to the temperature, film thickness and mismatch strain.

To minimize the electrostatic energy in Eq.(1), the equilibrium out-of-plane polarization $P_3$ can be inhomogeneous in both normal $z$ and transverse $x, y$ directions. Actually, the domain formation decreases the depolarization field energy, but increases the gradient-correlation energy. Their optimal balance determines the equilibrium distribution of $P_3$, e.g., z-profile of polarization, domain absence (for ultra-thin gaps) or domain period. In this **Section** we consider $P_3$ uniform in xy-plane, while the domains are considered in the next one. In a single domain state the polarization can change significantly only in an ultra-thin surface layer of angstrom order, at that in the case of natural boundary conditions, $\left.\frac{\partial P_3}{\partial z}\right|_{z=0,h} = 0$, consistent with Eq.(1), z-dependence of $P_3$ is typically absent [41, 42]. For a single-domain case, the average polarization $\bar{P}$ and dielectric susceptibility $\bar{\chi} = h\frac{d\bar{P}}{dV}$, as a functions of $V$, $T$, $h$ and $d$, satisfy the system of Landau-Khalatnikov decoupled equations:

$$\Gamma \frac{d}{dt}\bar{P} + \left(\alpha + \frac{d}{\varepsilon_0(\varepsilon_b d + \varepsilon_e h)}\right)\bar{P} + \beta\bar{P}^3 + \gamma\bar{P}^5 = \frac{\varepsilon_e V}{d\varepsilon_b + h\varepsilon_e}. \tag{2a}$$

$$\Gamma \frac{d}{dt}\bar{\chi} + \left(\alpha + \frac{d}{\varepsilon_0(\varepsilon_b d + \varepsilon_e h)} + 3\beta\bar{P}^2 + 5\gamma\bar{P}^4\right)\bar{\chi} = \frac{\varepsilon_e h}{d\varepsilon_b + h\varepsilon_e}. \tag{2b}$$

Here $\Gamma$ is a Khalatnikov coefficient, $\varepsilon_0$ is a universal dielectric constant, $\varepsilon_b \leq 10$ is a background permittivity of a ferroelectric [33] and $\varepsilon_e$ is a relative permittivity of the dielectric layer. The term $\frac{d}{\varepsilon_0(\varepsilon_b d + \varepsilon_e h)}$ originates from the depolarization field (see **Appendix A**). Note that the equilibrium susceptibility must be positive that yields $\left(\alpha + \frac{d}{\varepsilon_0(\varepsilon_b d + \varepsilon_e h)} + 3\beta\bar{P}^2 + 5\gamma\bar{P}^4\right) > 0$.

In **Appendix A**, we derive the static solution for the surface charge $Q$ at the electrodes and find the specific capacitance $C_{surf}$ per unit area as $dQ/dV$:

$$\frac{C_{surf}}{S} = \frac{\varepsilon_{eff}\varepsilon_0}{h+\varepsilon_b\frac{d}{\varepsilon_e}}. \tag{3a}$$

Here we introduced effective permittivity of the ferroelectric film

$$\varepsilon_{eff} = \frac{\varepsilon_b\varepsilon_0(\alpha+3\beta\bar{P}^2+5\gamma\bar{P}^4)+1}{\varepsilon_0\left(\alpha+\frac{d}{\varepsilon_0(\varepsilon_b d+\varepsilon_e h)}+3\beta\bar{P}^2+5\gamma\bar{P}^4\right)} \equiv \varepsilon_b + \frac{\varepsilon_e h}{\varepsilon_0(\alpha+3\beta\bar{P}^2+5\gamma\bar{P}^4)(\varepsilon_b d+\varepsilon_e h)+d}. \tag{3b}$$

It can be seen from the comparison of Eqs.(3a) and (3b), that the contributions of dielectric layer and ferroelectric films from (3a) cannot be separated, since the effective permittivity $\varepsilon_{eff}$ already contains the contribution from the dielectric layer. The numerical value of this contribution is close to $\varepsilon_b + \frac{\varepsilon_e h}{d}$ since $\varepsilon_0(\alpha + 3\beta \bar{P}^2 + 5\gamma \bar{P}^4) \cong 10^{-3}$. Hence expression (3b) principally differs from Salahudin et al. results [3]. This expression is qualitatively similar to results of Bratkovsky and Levanyuk [15], but differs quantitatively due to the consideration of background permittivity effect.

A usual approximation in analysis leads to the following expression for the capacitance:

$$\frac{C_{surf}}{S} \approx \frac{\varepsilon_0 \varepsilon_{\text{FE}}}{h + \varepsilon_{\text{FE}} \frac{d}{\varepsilon_e}}, \quad (3c)$$

where $\varepsilon_{\text{FE}}$ is the relative permittivity of a ferroelectric layer that is much higher than unity, $\varepsilon_{\text{FE}} \gg 1$. The numerators of Eqs.(3a) and (3c) contain different permittivities, $\varepsilon_{eff}$ and $\varepsilon_{\text{FE}}$. The denominators contain different dielectric constants, $\varepsilon_b$ and $\varepsilon_{\text{FE}}$, as well. The denominator of Eq.(3c) can be much bigger than those of Eq.(3a), if $\varepsilon_b + \frac{\varepsilon_e h}{d} \ll \varepsilon_{\text{FE}}^{eff}$. If it is so, the expression (3a) gives much higher values of capacitance then expected from the expression (3c).

It is important to note that the expression (3c) have been used by many authors to estimate the capacitance of the heterostructures. This model could explain "an enhancement" of dielectric response which can lead to "negative" values of some virtual dielectric susceptibility by many groups [6, 13]. In turn, this may account for the misinterpretations of experimental results with imperfect models like Eq.(3c).

Important, that $\varepsilon_{eff}$ can be only positive for equilibrium values of $\bar{P}$, which satisfy Eq.(2a), because the stability condition $\left(\alpha + \frac{d}{\varepsilon_0(\varepsilon_b d + \varepsilon_e h)} + 3\beta \bar{P}^2 + 5\gamma \bar{P}^4\right) > 0$ for the dielectric susceptibility strictly follows from Eq.(2b). This general conclusion is valid for any uniaxial ferroelectric film.

To illustrate this expected result, let us consider a model example prompted by the recent interest to 2D ferroelectrics, namely the temperature and size dependences of the capacitance (3a) for the layered ferroelectric $Sn_2P_2S_6$ (**SPS**). Its thermodynamic parameters are summarized in the **Table II.** The dielectric permittivity of the dielectric layer, $\varepsilon_e = 25$, corresponds to e.g., a bulk $HfO_2$ or paraelectric oxide.

**Table II.** The parameters for bulk ferroelectric $Sn_2P_2S_6$ collected from Refs.[43, 44]

| parameter | dimension | value |
|---|---|---|
| $\varepsilon_b^*$ | dimensionless | 7 |

| $\alpha_T$ | m/F | $1.44 \times 10^6$ |
|---|---|---|
| $T_C$ | K | 337 |
| $\beta$ | $C^{-4} \cdot m^5 J$ | $9.40 \times 10^8$ |
| $\gamma$ | $C^{-6} \cdot m^9 J$ | $5.11 \times 10^{10}$ |
| $g_{11}$** | $m^3/F$ | $5.0 \times 10^{-10}$ |
| $g_{44}$** | $m^3/F$ | $2.0 \times 10^{-10}$ |

\* estimated from a refraction index value

\*\* typical values, which order of magnitude is estimated from the uncharged domain wall width [43, 44]

The temperature dependance of the SPS film static capacitance calculated for several values of the film thickness $h =$ (5 - 20) nm is shown in **Fig. 3**. The capacitance of the dielectric layer is very small and positive, around $10^{-22}$ F/m$^2$ (see horizontal line). The capacitance of the ferroelectric film is also positive, and its divergency corresponds to the thickness-dependent temperature of the ferroelectric-paraelectric phase transition.

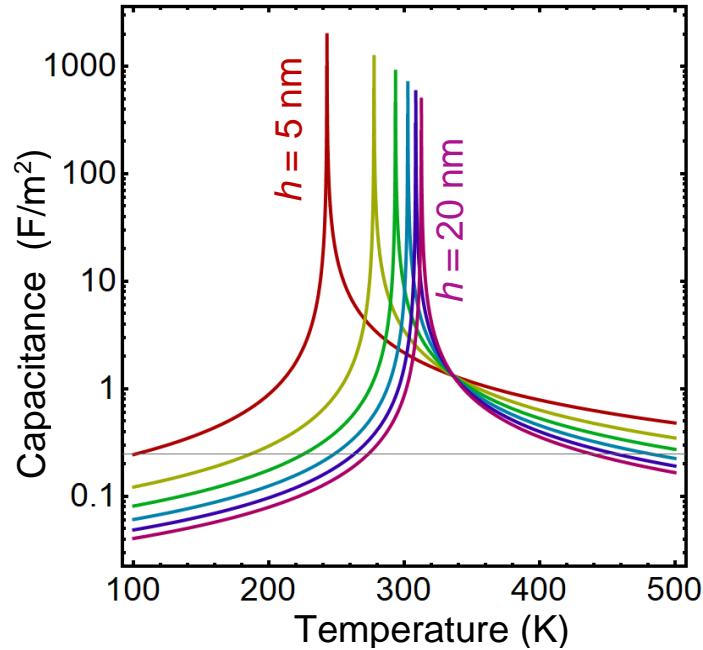

**Figure 3.** The temperature dependance of the SPS layer steady-state capacitance calculated for several values of the film thickness $h = 5$ nm (red curve), 8 nm (yellow curve), 11 nm (green curve), 14 nm (light blue curve), 17 nm (dark blue curve) and 20 nm (purple curve). The dielectric layer thickness $d =$ 0.2 nm.

The dependencies of the polarization, capacity, and normalized dielectric susceptibility of the SPS film on the electric field $E$ applied to the system were calculated for several temperatures $T = (50 – 300)$ K and are shown in **Fig. 4**. Here the dimensionless frequency is in the units of Khalatnikov relaxation time $\tau_{Kh} = \Gamma/|\alpha_T T_c|$.

**Figs. 4(d-f)**, calculated from Eqs.(2) for a very small external field frequency $\omega\tau_{Kh} = 10^{-4}$, contain the conventional square $P(E)$ hysteresis loops, positive capacity and dielectric susceptibility with very sharp peaks at coercive field. **Figures 4(a-c)**, calculated for a higher frequency $\omega\tau_{Kh} = 5 \cdot 10^{-2}$, illustrate the effect of the hysteresis loops smearing and blowing, in complex with diffusing of capacity and dielectric permittivity maxima at coercive field. Note that the capacitance and the susceptibility have very similar $E$-dependences, but differ in the order of magnitude.

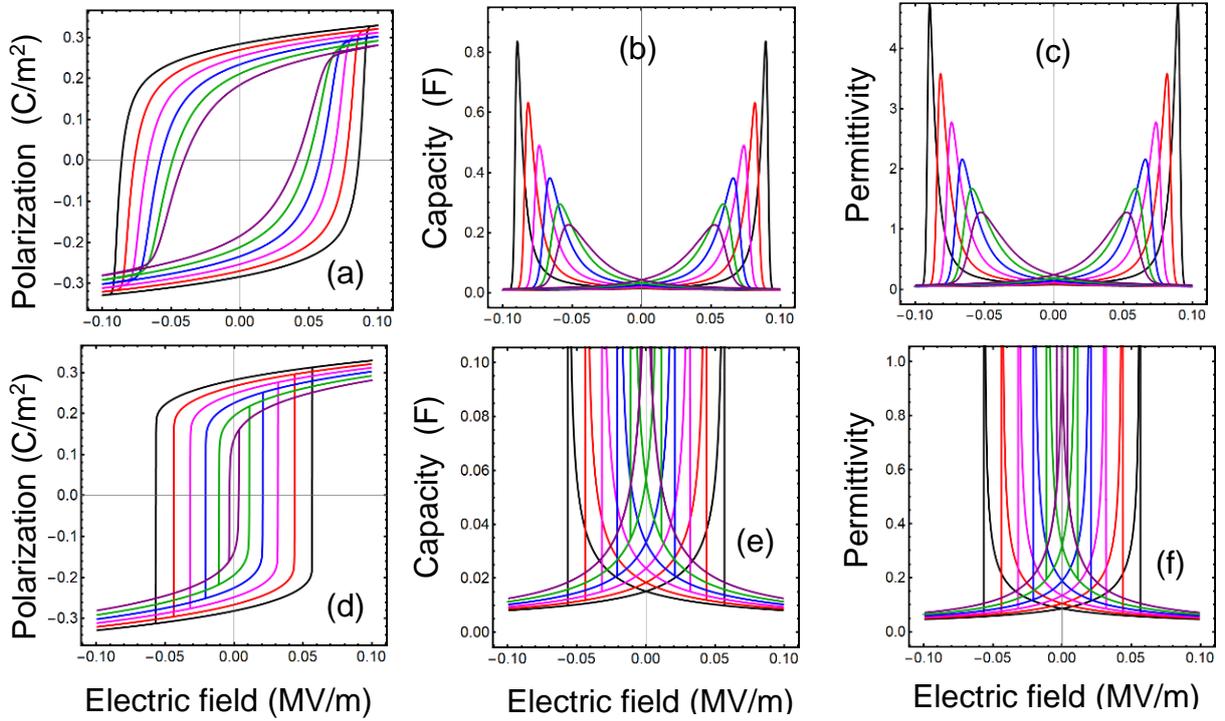

**Figure 4**. The dependencies of the polarization **(a, d)**, the capacitance **(b, e)**, and the susceptibility **(c, f)** of the 50-nm thick SPS film on the applied electric field calculated for several temperatures $T = 50$ K (black curves), 100 K (red curves), 150 K (magenta curves), 200 K (blue curves), 250 K (green curves) and 300 K (purple curves). The bottom row is for a very small external field frequency $\omega\tau_{Kh} = 10^{-4}$, and the top one is for a higher frequency $\omega\tau_{Kh} = 5 \cdot 10^{-2}$. The dielectric layer thickness $d = 0.2$ nm, the Khalatnikov time $\tau_{Kh} = \Gamma/|\alpha_T T_c|$.

From **Fig. 4** the capacitance is positive in both the static and dynamic cases, both for high and low frequencies. The result for a static case following from the well-known thermodynamic theorem.

We note that multiple groups including Salahuddin and Datta [13], have found that under certain conditions the electrostatic energy, associated with dielectric layer and depolarization field inside the ferroelectric film, could suppress two wells (corresponding to $P = \pm P_s$) of intrinsic LGD free energy functional. Keeping this fact in mind, it could be said that the intrinsic free energy of a ferroelectric should be used in order to get the *local* susceptibility of a ferroelectric layer in a multilayer system and obtain some a negative susceptibility in this case. At the same time the long-range electrostatic interaction between the two layers caused by the depolarization field significantly complicates the division of the total free energy on intrinsic "ferroelectric" and extrinsic "dielectric" parts.

### III. Capacitance Under the Presence of a Domain Structure

We further explore the potential of the NC emergence in the presence of domain structures. Following recent review of Hoffmann et al. [6], it is important to distinguish intrinsic and extrinsic effects of NC. The latter determined by the presence of ferroelectric domains, while former is related to alleged NC effects of single domain ferroelectrics.

The stability of the paraelectric phase with respect to the formation of a periodic domain structure was studied analytically and high-accuracy expressions near the paraelectric-ferroelectric phase boundary were derived in details in **Appendix B**. Using direct variational method for the LGD free energy (1), we introduce the polarization trial function

$$P_3(x,z) \approx P + 2A\cos(kx)\left[\cosh(q_1(z-h)) - \frac{q_1}{q_2}\sinh(q_1 h)\exp(-q_2 z)\right], \quad (4a)$$

where the polarization magnitude $P$, amplitude $A$ and wave vector $k$ are variational parameters; and the increments $q_1$ and $q_2$ are expressed though $k$ and other parameters listed in Table I as

$$q_1 \approx k\sqrt{\frac{\alpha + g_{44}k^2}{\alpha + g_{44}k^2 + g_{11}k^2 + \frac{1}{\varepsilon_0\varepsilon_b}}}, \qquad q_2 \approx \sqrt{\frac{1}{\varepsilon_0\varepsilon_b g_{11}} - q_1^2}. \quad (4b)$$

Note, that the trial function (4a) is built to satisfy natural boundary conditions. Consequently, the derived expressions (4b), as well as all further expressions, are much more rigorous in comparison with earlier papers (see e.g., [45]). Using this function, we obtain the free energy

$$\Delta G \approx \frac{\alpha_p}{2}P^2 + \frac{\alpha_a}{2}A^2 + \frac{\beta}{4}\left(P^4 + 6P^2 A^2 + \frac{9}{4}A^4\right) - E_{eff}P \quad (5a)$$

with renormalized coefficients and effective electric field, which are given by expressions:

$$\alpha_p = \alpha + \frac{d}{\varepsilon_0(d\varepsilon_b + h\varepsilon_e)}, \quad \alpha_a = \alpha - g_{11}q_1^2 + g_{44}k^2 - \frac{1}{\varepsilon_0\varepsilon_b}\frac{q_1^2}{k^2 - q_1^2}, \quad E_{eff} = \frac{\varepsilon_e}{d\varepsilon_b + h\varepsilon_e}V. \quad (6a)$$

The magnitude $P$ and the amplitude $A$ satisfy the equations

$$A = \pm \frac{2}{3}\sqrt{\frac{-\alpha_a - 3\beta P^2}{\beta}}, \qquad \left(\alpha_p - \frac{4}{3}\alpha_a\right)P - 3\beta P^3 = E_{eff}. \tag{6b}$$

Due to the negative sign before the second term in Eq.(6b) for $P$, it leads to a "mirrored" hysteresis curve. The solution is stable when

$$-2(\alpha_a + 3\beta P^2) > 0 \quad \text{and} \quad \alpha_p - \frac{4}{3}\alpha_a - 9\beta P^2 > 0. \tag{7}$$

The measurable (i.e., average) dielectric susceptibility in the presence of periodic domain structure is given by expression:

$$\frac{\partial \langle P_3 \rangle}{\partial E_{eff}} = \frac{\partial P}{\partial E_{eff}} = \frac{1}{\alpha_p - \frac{4}{3}\alpha_a - 9\beta P^2}. \tag{8}$$

As anticipated, the static susceptibility (8) is always positive according to the stability conditions (7). This excludes the viability of the steady-state NC.

However, rather high electric fields could make the single-domain ferroelectric (**SDFE**) phase stable, and poly-domain ferroelectric (**PDFE**) phase unstable with a negative permittivity. At the same time poly-domain states manifest itself under the quasi-static electric field by double hysteresis loops of antiferroelectric (AFE) type, which are characterized by the first and the second critical electric fields. The schematics of the AFE-type bistable curves is shown in **Fig. 1(e).**

To obtain the first critical field, we rewrite the second inequality in Eq.(6b) as the upper limit for polarization $9\beta P^2 < \alpha_p - 4\alpha_a/3$, and obtain

$$E_{cr}^{(PDFE)} = \frac{2}{3}\left(\alpha_p - \frac{4}{3}\alpha_a\right)\sqrt{\frac{1}{9\beta}\left(\alpha_p - \frac{4}{3}\alpha_a\right)}. \tag{9a}$$

The PDFE phase is stable for the fields $|E_{eff}| < E_{cr}^{(PDFE)}$. The second critical field can be obtained from the condition of zero amplitude $A$ in the first of Eq.(6b) achieved at $P = \sqrt{-\alpha_a/3\beta}$, which corresponds to the expression

$$E_{cr}^{(SDFE)} = \left(\alpha_p - \frac{\alpha_a}{3}\right)\sqrt{\frac{-\alpha_a}{3\beta}}. \tag{9b}$$

The SDFE phase is stable for the fields $|E_{eff}| > E_{cr}^{(SDFE)}$.

The above expressions depend on the wave vector $k$ of the domain structure, which equilibrium value, $k_{eq}$, has the form:

$$k_{eq} = \begin{cases} 0, & 0 \leq d < d_{cr}, \\ \sqrt{\frac{1}{h}\sqrt{\frac{2}{\varepsilon_0 \varepsilon_b g_{44}}} - \frac{2}{h^2}\left(1 + \frac{\varepsilon_e h}{\varepsilon_b d}\right)}, & d_{cr} \leq d \leq 10 d_{cr}, \\ \sqrt{\frac{\pi}{2h\sqrt{\varepsilon_0 \varepsilon_b g_{44}}} - \frac{\pi^2}{4h^2} - 2\frac{\varepsilon_e}{\varepsilon_b h\, d}}, & d > 10 d_{cr}. \end{cases} \tag{10a}$$

Here $d_{cr}$ is the critical thickness of the dielectric layer above which the domain splitting starts. The approximate expression for $d_{cr}$ is:

$$d_{cr} = \frac{\varepsilon_e}{\varepsilon_b}\sqrt{2\varepsilon_0 \varepsilon_b g_{44}}. \tag{10b}$$

The condition $d < d_{cr}$ is required for the absolute stability of a single-domain state. For thin dielectric layers ($d < d_{cr}$) the domain structure is unstable and the paraelectric phase transforms into a single domain ferroelectric phase upon cooling. The condition $d \gtrsim d_{cr}$ corresponds to the appearance of the very broad domains with small $k_{eq}$. The "thick" dielectric layer $d \gg d_{cr}$ induces Kittel-like domains with the width $\sim\sqrt{h}$.

Note that Eqs.(10) are derived within the LGD approach, which accounts self-consistently for the domain wall finite width and broadening effect via the polarization gradient coefficient $g_{44}$ and for the contribution of depolarization field proportional to $\varepsilon_0 \varepsilon_b$. In Ref. [15] the Kittel model of infinitely thin domain wall was used. While the technique for analytical consideration of the domain formation threshold was published long ago [20, 22], the qualitative physical picture and corresponding analytical expressions (4)-(10) are derived in this work. The important question is how accurate are these expressions?

We emphasize that the Eqs.(4)-(10) are exact at the onset of the domain formation, and, at the same time, they are valid for any uniaxial ferroelectric film, being non-material specific. Their accuracy becomes lower when the domain structure profile acquires anharmonic features. The ferroelectric multiaxiality can principally change the domain morphology due to the polarization rotation making the equilibrium domain configuration much more complex and versatile. Specifically, the flux-closure domains appear in the film near the boundary with the dielectric layer [46, 47], and it is not excluded that a complex, e.g., vortex-type, domain topology in e.g., ferroelectric/paraelectric superlattices, can be a reason of NC effect at the domain onset.

Next, we performed the finite element modeling (**FEM**) of the 2D boundary problem [geometry is shown in **Fig. 2(a)**] consisting in the dynamic LGD-type differential equations coupled with the Poisson equation, which are obtained by the variation of the free energy (1) allowing for the Khalatnikov relaxation of polarization (see **Appendix C**).

The critical thickness $d_{cr}$ calculated from Eq.(10b) is equal to 0.6 nm for a SPS film, and the value perfectly agrees with FEM results. Actually, a 50-nm thick film relaxes to a homogeneously polarized state at $d <0.6$ nm. For bigger $d$ FEM results, shown in **Fig. 5**, illustrate the changes of the equilibrium domain structure near the critical thickness of the dielectric layer. For $d =0.8$ nm domain stripes are relatively wide corresponding to the domain onset, and a weak broadening of the domain walls, originated from the depolarization field effect [21], is seen near the dielectric layer [**Fig. 5(a)**]. For $d$ =0.9 nm the domain period, $2\pi/k_{eq}$, sharply decreases and the broadening effect becomes more pronounced [**Fig. 5(b)**]. For $d$ =1 nm and 1.2 nm the domain splitting continues, equilibrium domains become significantly thinner, their walls become thicker and much more broaden approaching the dielectric layer, however the ferroelectric-paraelectric transition does not occur for a 50 nm film with the further increase of $d$ [compare **Figs. 5(c)** and **5(d)**].

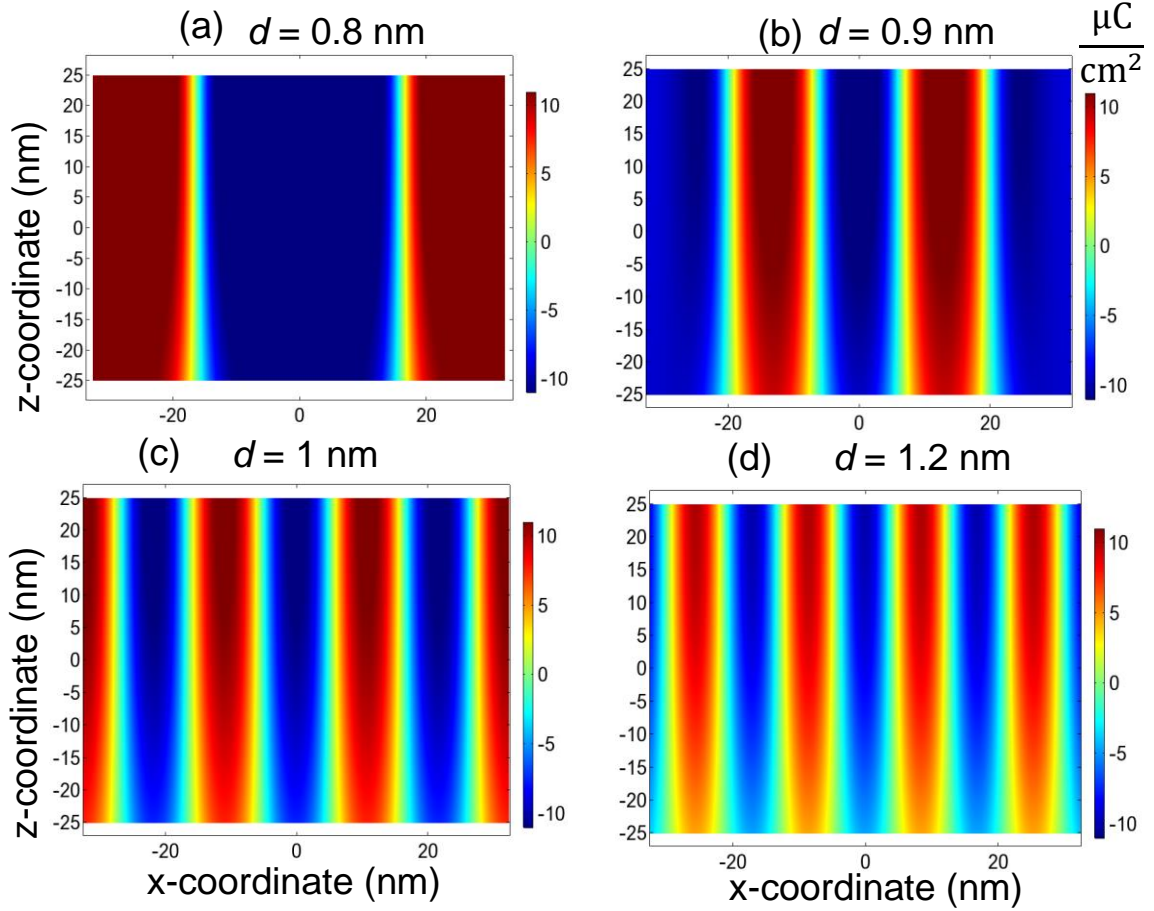

**Figure 5**. Polarization distribution in the vertical cross-section of the system SPS film /dielectric layer for several values of the dielectric layer thickness $d$=0.8 nm **(a)**, 0.9 nm **(b)**, 1 nm **(c)** and 1.2 nm **(d)**. The film thickness $h$ =50 nm, $T = 300$ K. The equilibrium domain structure is shown.

Note that $d_{cr}$ appeared independent on the film thickness and temperature in accordance with Eq.(10b). This result is the consequence of a sinusoidal domain profile at their onset. The thickness dependence of the domain period ($2\pi/k_{eq}$) near the transition to paraelectric phase is shown in **Fig. 6(a)** for different values of the dielectric layer thickness $d = (0.6 - 10)$nm. It is seen that the period exists for all $d > d_{cr}$ (red, brown and black curves); and even for $d = d_{cr}$ and $h > 5$ nm (blue curve). However, this result is formal, because the conditions of the ferroelectric phase stability are not super-imposed on the conditions (10).

Allowing for the temperature dependence of the coefficient $\alpha = \alpha_T(T - T_c)$, the boundaries between the single-domain, domain onset and poly-domain states correspond to the conditions:

$$\begin{cases} \alpha_T(T - T_c) + \dfrac{d}{\varepsilon_0(\varepsilon_b d + \varepsilon_e h)} = 0, & 0 \leq d < d_{cr}, \\ \alpha_T(T - T_c) + g_{44}\left[\sqrt{\dfrac{2}{\varepsilon_0 \varepsilon_b g_{44}}} - \left(\dfrac{1}{h} + \dfrac{\varepsilon_e}{\varepsilon_b d}\right)\right]\dfrac{2}{h} = 0, & d_{cr} \leq d \leq 10 d_{cr}, \\ \alpha_T(T - T_c) + g_{44}\left(\dfrac{\pi}{\sqrt{\varepsilon_0 \varepsilon_b g_{44}}} - \dfrac{\pi^2}{4h} - 2\dfrac{\varepsilon_e}{\varepsilon_b d}\right)\dfrac{1}{h} = 0, & d > 10 d_{cr}. \end{cases} \quad (11)$$

Phase diagrams of SPS film, shown in **Fig. 6(b-d)**, contain three stable phases, namely, the paraelectric (**PE**), the ferroelectric single-domain (**SDFE**) and ferroelectric polydomain (**PDFE**) phases. The transition temperature as a function of $d$ calculated for different $h$ values is shown in **Fig. 6(b)**. The SDFE is stable for thin dielectric layers and borders with PE phase; but the phase does not exist for thick dielectric layers (solid curves), while the PDFE does (dashed curves). The region of FE phases increases with $h$ increase, as anticipated. The PE-FE transition temperatures as a function of $h$ is shown in **Fig. 6(c)** for $d = 0.6$ nm and in **Fig. 6(d)** for $d = 3$ nm. Since $d_{cr} \approx 0.6$ nm, only the PE and SDFE phases exist in **Fig. 6(c)**; while three phases PE followed by PDFE and then by SDFE exist in **Fig. 6(d)**.

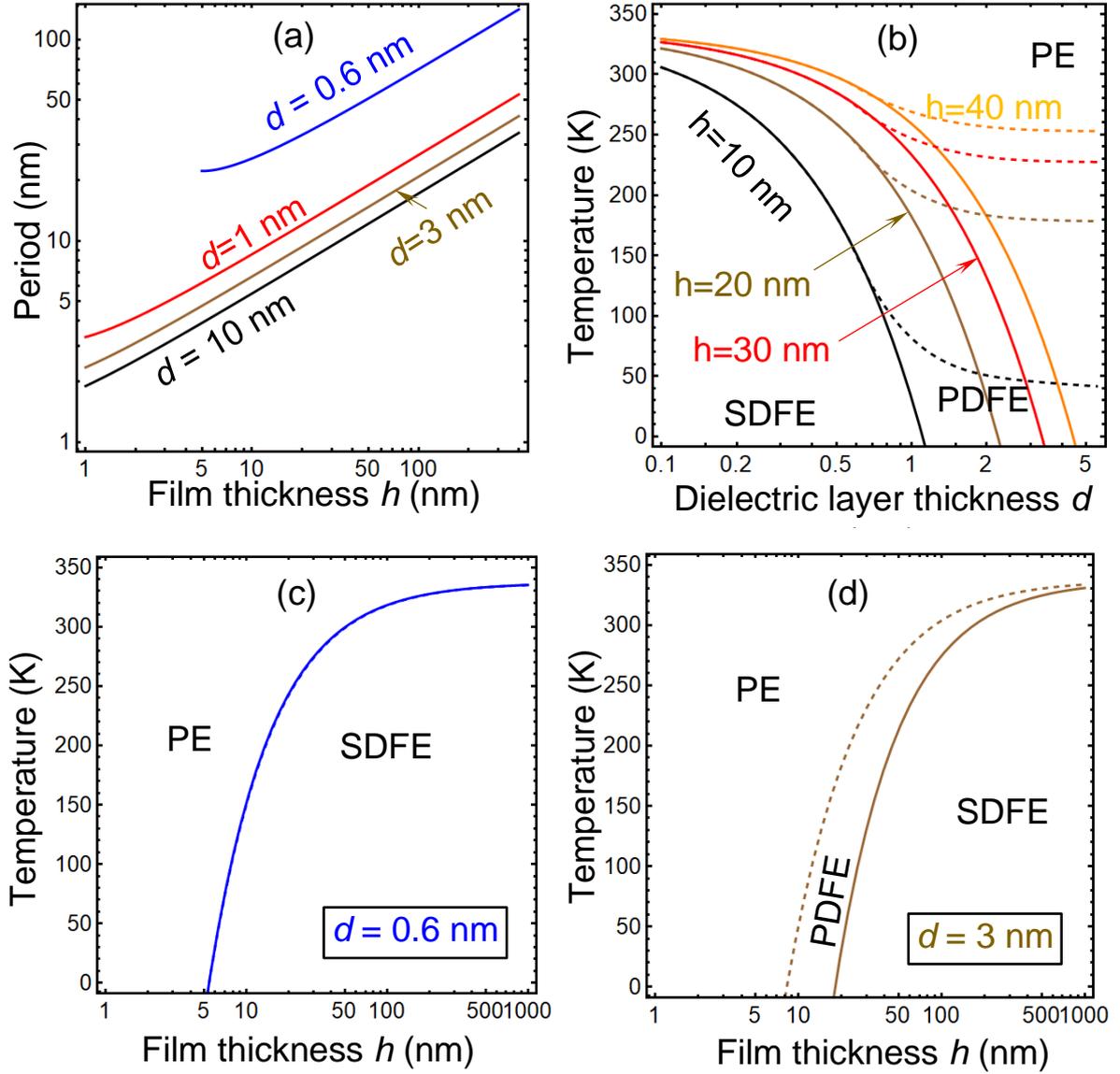

**Figure 6**. (a) The dependence of the equilibrium domain period ($2\pi/k_{eq}$) on the SPS film thickness $h$ calculated near the transition to the paraelectric (PE) phase for different values of the dielectric layer thickness $d$ (shown near the curves). (b) The transition temperature as a function of $d$ calculated for different $h$ values (shown near the curves). (c, d) The PE-FE transition temperatures as a function of $h$ calculated for $d = 0.6$ nm (c) and $d = 3$ nm (d). Solid and dashed curves represent the stability boundaries of the single domain (SDFE) and polydomain (PDFE) ferroelectric phases, respectively.

Also, we performed FEM of the quasi-static polarization and dielectric susceptibility hysteresis loops of the SPS film covered by a dielectric layer of thickness close to the critical $d \approx d_{cr}$, and by a thicker layer $d > d_{cr}$. We obtained very slim loops of polarization for $d \approx d_{cr}$, which quasi-elliptic shape at low voltages acquires super-linear endings with voltage amplitude increasing [compare black, red, green and blue curves in **Fig. 7(a)**]. Corresponding loops of quasi-static susceptibility are positive and acquires a pronounced U-like shape under voltage amplitude increasing [see different curves in **Fig. 7(b)**]. For higher $d > d_{cr}$ polarization loops are either almost hysteresis less at small voltages, or antiferroelectric-type shape with voltage amplitude increasing [compare black, red, green and blue curves in **Fig. 7(c)**]. The loops of quasi-static susceptibility have a U-like shape at low voltages, and become bow-like with voltage amplitude increasing [compare different curves in **Fig. 7(d)**].

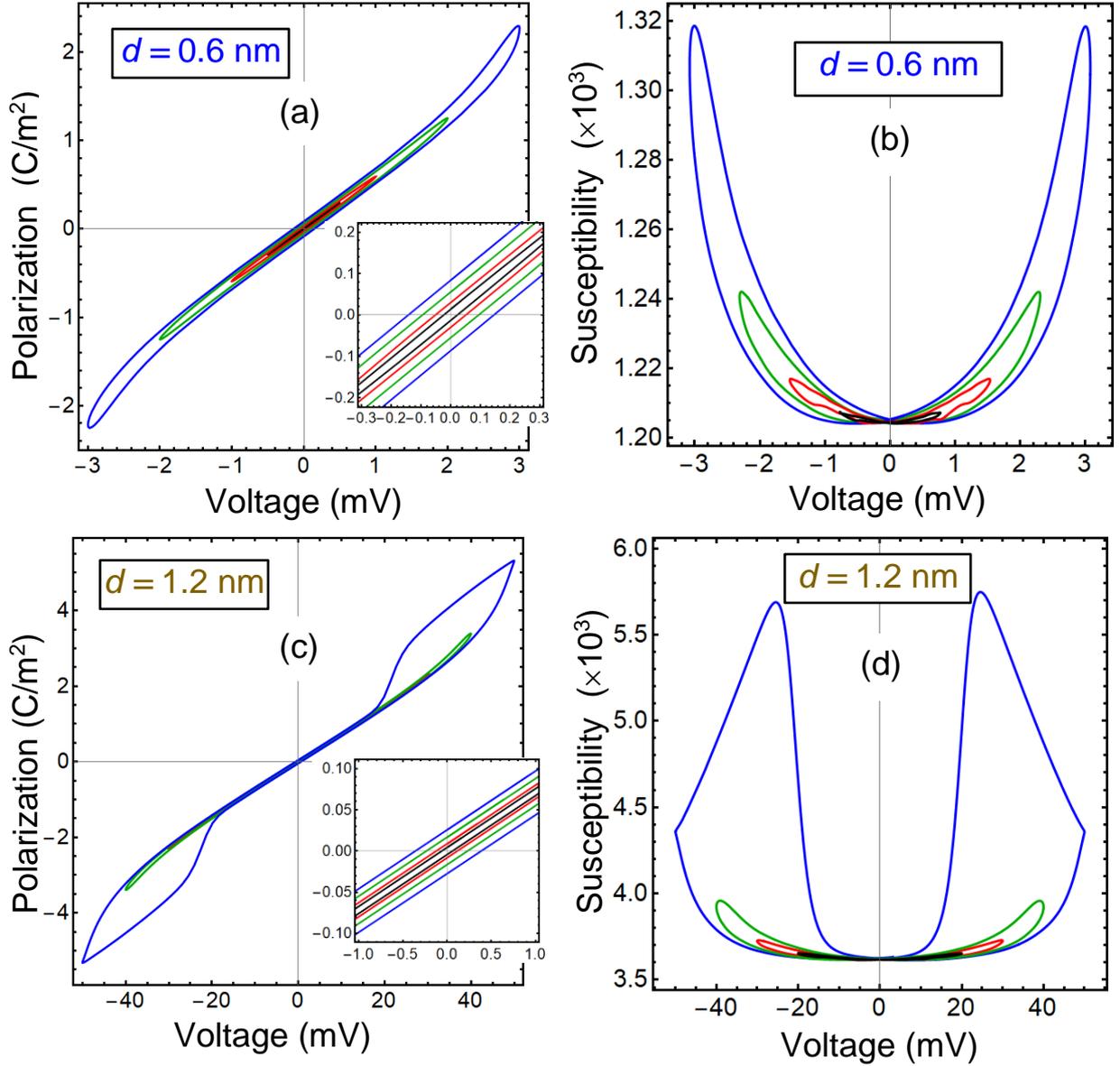

**Figure 7**. The quasi-static dependencies of the polarization **(a, c)** and the dielectric susceptibility **(b, d)** of the SPS film on the applied voltage calculated for several values of the dielectric layer thickness $d$=0.6 nm **(a, b)** and 1.2 nm **(c, d)**. The voltage amplitude is $V = 0.5$ mV (black curves), 1 mV (red curves), 2 mV (green curves) and 3 mV (blue curves) for plots **(a-b)**; and $V = 10$ mV (black curves), 30 mV (red curves), 40 mV (green curves) and 50 mV (blue curves) for plots **(c-d)**. The film thickness $h$ =50 nm, $T = 300$ K.

**IV. Capacitance Under the Presence of a Sluggish Nonlinear Surface Screening**

Building upon the previous analysis, we consider the system consisting of an electron-conducting bottom electrode, an SPS film covered with a layer of surface charge (e.g., ions, electrons, protons, OH⁻, and/or holes), which charge density $\sigma(\phi)$ depends on the acting electric potential $\phi$ in a complex nonlinear way. An ultra-thin dielectric gap separates the film surface and the top electrode, that is either ion-conductive planar electrode or flatted apex of SPM tip. The electrode provides a direct ion exchange with an ambient media, as shown in **Fig. 2(b).**

To describe the dynamics of the surface charge density, we use a linear relaxation model [48, 49] for positive ($\sigma_p$) and negative ($\sigma_n$) charges:

$$\tau_p \frac{\partial \sigma_p}{\partial t} + \sigma_p = \sigma_{p0}[\phi], \quad \tau_n \frac{\partial \sigma_n}{\partial t} + \sigma_n = \sigma_{n0}[\phi]. \tag{12a}$$

where the relaxation times $\tau_p$ and $\tau_p$ can be very different due to different mobility of positive and negative charges. The total surface charge is $\sigma = \sigma_p + \sigma_n$.

The dependence of equilibrium charge densities, $\sigma_{p0}[\phi]$ and $\sigma_{n0}[\phi]$, on the electric potential $\phi$ can be defined in various ways. We assume the way controlled by their concentration at the interface $z = 0$ in a self-consistent manner, e.g., as proposed by Stephenson and Highland (**SH**) [50, 51]. In this work we will use a natural extension of the SH model:

$$\sigma_{p0}[\phi] = \frac{eZ_p}{A_p} \left(1 + g_p \exp\left(\frac{\Delta G_p^0 + eZ_p \phi}{k_B T}\right)\right)^{-1}, \tag{12b}$$

$$\sigma_{n0}[\phi] = \frac{eZ_n}{A_n} \left(1 + g_n \exp\left(\frac{\Delta G_n^0 + eZ_n \phi}{k_B T}\right)\right)^{-1}, \tag{12c}$$

where $e$ is an elementary charge, $Z_{p,n}$ is the ionization degree of the surface charges, $1/A_{p,n}$ are saturation densities of the charges. Positive parameters $\Delta G_p^0$ and $\Delta G_n^0$ are the free energies of the surface defects formation at normal conditions and zero potential $\phi = 0$. Exact values of $\Delta G_{p,n}^0$ are poorly known for many practically important cases. Positive prefactors $g_p$ and $g_n$ can originate from different mechanics of the charge formation [52, 53]. The condition of charge absence should be valid for the sum $\sigma_{p0}[0] + \sigma_{n0}[0]$ in the thermodynamic equilibrium, and the condition imposes limitations on the charge density parameters in Eqs.(12).

Far from the equilibrium, e.g., during the electrochemical reactions, the sum $\sigma_{p0} + \sigma_{n0}$ can be nonzero even at $\phi = 0$. The latter case was considered by SH, who propose that $g_p = \rho^{1/N_p}$ and $g_n = \rho^{1/N_p}$, where $\rho = \frac{p_{O_2}}{p_{O_2}^{00}}$ is the relative partial pressure of an ambient gas (e.g., oxygen), which varied in a wide range from $10^{-6}$ to $10^6$ [50, 51]; and $N_{n,p}$ is the number of positive and negative surface ions created per gas molecule.

To analyze a steady-state capacitance, we impose the condition $\sigma_{p0} + \sigma_{n0} = 0$. For the purpose we regard that the ratios $\frac{eZ_p}{A_p}$ and $\frac{eZ_n}{A_n}$ are opposite, i.e., $\frac{eZ_p}{A_p} = -\frac{eZ_n}{A_n}$, the formation energies are equal $\Delta G_p^0 = \Delta G_n^0$ and vary in the range ~(0.03 – 0.3) eV [50]; and either $g_p = g_n \approx 1$ or $g_p = g_p = \rho^{1/N_p}$ along with the condition $N_p = N_n$. The violation of these conditions, e.g., $N_p = -N_n$ were considered elsewhere [50, 51].

Since the stabilization of a single-domain polarization in ultrathin perovskite films covered by surface ions typically takes place (see e.g. Refs.[54, 55, 56]), we will assume that the distribution of the out-plane component $P_3(x,z)$ does not deviate significantly from the polarization value averaged over the film thickness, namely $P_3 \cong \bar{P}$. In a single domain state, the Landau-Khalatnikov equation, determining the average polarization $\bar{P}$ as a function of $V, T, h$ and $d$, has the form [49]:

$$\Gamma \frac{d}{dt}\bar{P} + \alpha_T (T - T_c)\bar{P} + \beta \bar{P}^3 + \gamma \bar{P}^5 = \frac{\Psi}{h}. \tag{13a}$$

Here overpotential $\Psi$ contains the contribution from the surface charge density $\sigma_p + \sigma_n$, the depolarization field contribution proportional to $\bar{P}$, and the external potential drop proportional to the applied voltage $V$. Allowing for results in Ref.[49] and Eqs.(12), $\Psi$ can be determined in a self-consistent manner:

$$\frac{\Psi}{h} = \frac{d}{\varepsilon_0(\varepsilon_e h + d\varepsilon_b)}(\sigma_p + \sigma_n - \bar{P}) + \frac{\varepsilon_e V}{\varepsilon_e h + d\varepsilon_b}, \tag{13b}$$

$$\tau_p \frac{\partial \sigma_p}{\partial t} + \sigma_p = \frac{eZ_p}{A_p}\left(1 + g_p \exp\left(\frac{\Delta G_p^0 + eZ_p \Psi}{k_B T}\right)\right)^{-1}, \tag{13c}$$

$$\tau_n \frac{\partial \sigma_n}{\partial t} + \sigma_n = \frac{eZ_n}{A_n}\left(1 + g_n \exp\left(\frac{\Delta G_n^0 + eZ_n \Psi}{k_B T}\right)\right)^{-1}. \tag{13d}$$

Hence, we obtained a coupled system of four nonlinear equations (13), with three different characteristic time, $\Gamma$, $\tau_p$ and $\tau_n$. These times can be very different, and also different from the period of applied voltage $\tau_V$.

Typically, the relaxation of electrons is faster than that for the holes, and much faster than the relaxation of ions, thus the inequalities $\tau_n \leq \tau_p$ or even $\tau_n \ll \tau_p$ are valid. Since a polarization relaxation is determined by soft optic phonons, the strong inequality, $\Gamma/|\alpha| \ll \tau_{n,p}$, is valid far from the Curie temperature; and it makes sense to normalize time $t$ in Eqs.(13) on the Khalatnikov time, $\tau_{Kh} = \Gamma/|\alpha_T T_c|$. Note that the screening charge $\sigma(V)$ is located at the film-gap interface, the charge of the bottom electrode is $\sigma_{be} = \frac{\varepsilon_e h}{\varepsilon_e h + \varepsilon_b d}(\bar{P} - \sigma)$, and the charge of the top electrode is $\sigma_{te} = -\frac{\varepsilon_e h}{\varepsilon_e h + \varepsilon_b d}\bar{P} - \frac{\varepsilon_b d}{\varepsilon_e h + \varepsilon_b d}\sigma$.

Since the system of the coupled Eqs.(13) does not allow for analytical solutions, we studied the numerical solution for polarization and screening charges in a wide range of film thickness (4

– 200) nm, temperatures (50 – 300) K, formation energies $\Delta G_{p,n}^0 = (0.03 - 0.3)$ eV, and relaxation times $10 \leq \tau_n \leq 10^3$. Typical voltage dependences $\bar{P}(V)$, $\sigma_n(V)$, $\sigma_p(V)$, and $\sigma_{be}(V)$, their derivatives $\frac{\partial \bar{P}}{\partial V}$ and $\frac{\partial \sigma_{be}}{\partial V}$ are shown in **Fig. 8**. They are calculated for a 100-nm SPS film, room temperature and $\tau_n = \tau_p$. The same dependences shown in **Fig. 9** are calculated for low temperature and $\tau_n \leq \tau_p$. The frequency $w$ of applied voltage is very low, $w\tau_{Kh} \leq 10^{-3}$, in both figures, where we show the stationary loops only and omit transient processes. Other parameters are: $A_n = A_p = A = 10^{-18}$ m², $Z_p = -Z_n = +1$, $g_p = g_n = 1$, and $\Delta G_p^0 = \Delta G_n^0 = 0.1$ eV.

The quasi-static dependencies of the polarization $\bar{P}(V)$ do not have any unstable regions with negative slope, because the electric cycling (even with very low frequency) prevents the system following the unstable part of the $\bar{P}(V)$ curve [see black curves in **Figs. 8(a), (c), (e)** and **9(a), (c), (e)**]. The voltage behavior of negative and positive screening charges, $\sigma_n(V)$ and $\sigma_p(V)$, are inverted in the case $\tau_n = \tau_p$ when $\sigma_n(V) = \sigma_p(-V)$ [compare blue and red curves in **Figs. 8(a), (c), (e)** and **9(a)**]. The inversion disappears in the case $\tau_n \ll \tau_p$ [compare blue and red curves in **Figs. 9(c)** and **(e)**].

If the relaxation of at least one type of the screening charges is very slow in comparison with polarization response to external electric field, i.e., $\tau_{n,p} \gg \tau_{Kh}$, the strong retardation between the polarization bound charges and the screening charges appear during the quasi-static electric field cycling. The retardation effects manifest in the unusual features of the hysteresis loops $\sigma_{be}(V)$ [see green curves inside the dotted circle in **Figs. 8(a), (c), (e)** and **9(a), (c), (e)**]. The features look like two antisymmetric peaks ("up" and "down") located near the coercive voltage points of polarization. The derivative of electrode charge $\sigma_{be}$ with respect to the applied voltage, $\frac{\partial \sigma_{be}}{\partial V}$, can be negative in the region of the peaks [see negative parts of green curves in **Figs. 8(b), (d), (f)** and **9(b), (d), (f)**].

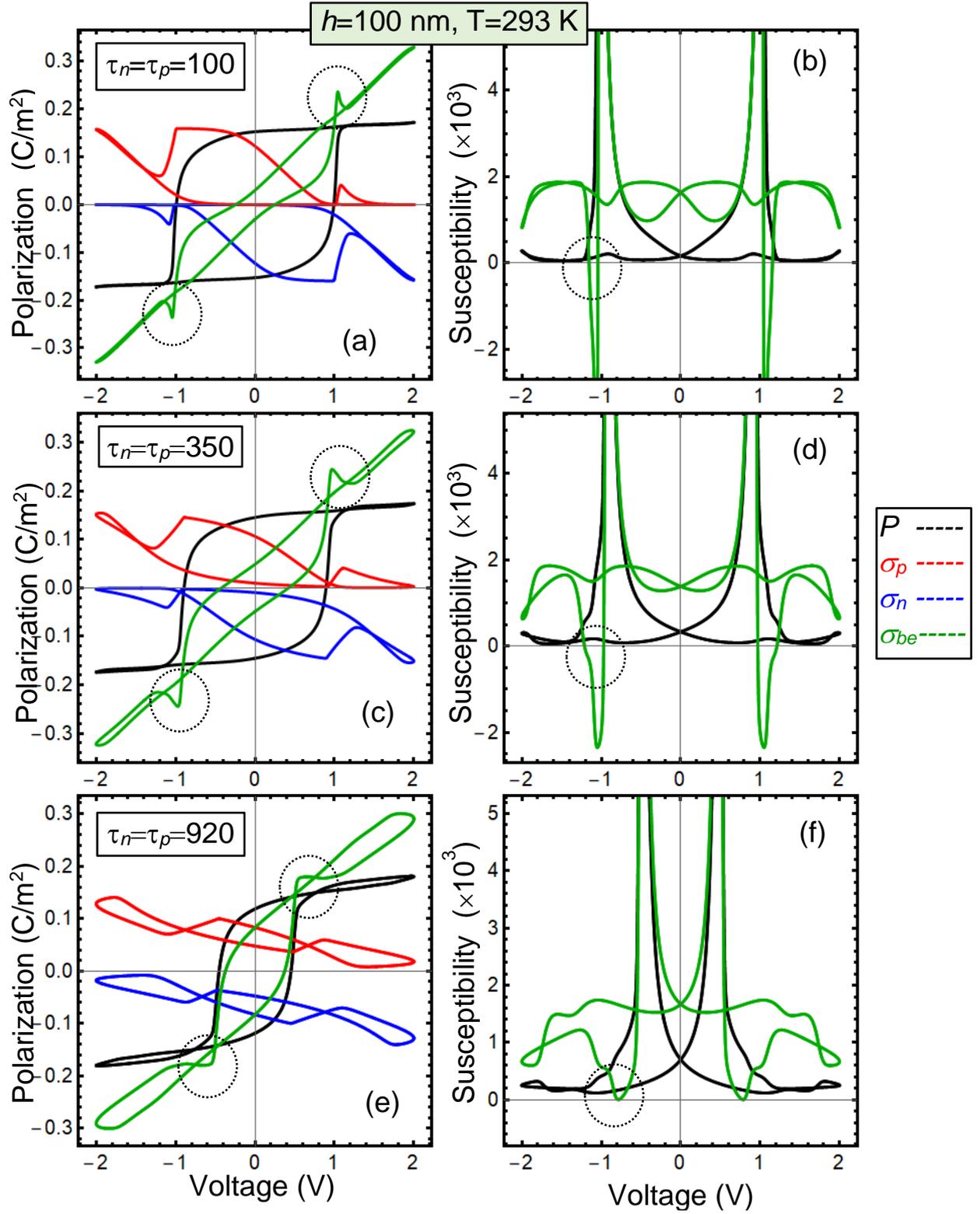

**Figure 8**. **(a, c, e)** The quasi-static dependencies of the polarization (black curves), positive (red curves) and negatives (blue curves) surface charges, and the total charge at bottom electrode (green curves). **(b, d, f)** The quasi-static dependencies of the dielectric susceptibility (black curves) and effective capacitance (green curves). The dependences are calculated for several values of characteristic times $\tau_n = \tau_p = 100$ **(a,b)**, $\tau_n = \tau_p = 350$ **(c, d)**, and $\tau_n = \tau_p = 920$ **(e, f)** in the units of $\tau_{Kh}$. The SPS film thickness $h = 100$ nm,

dielectric gap thickness $d = 1.2$ nm, $T = 293$ K. The quasi-static voltage with an amplitude is 2V and dimensionless very low frequency $10^{-3}$ is applied.

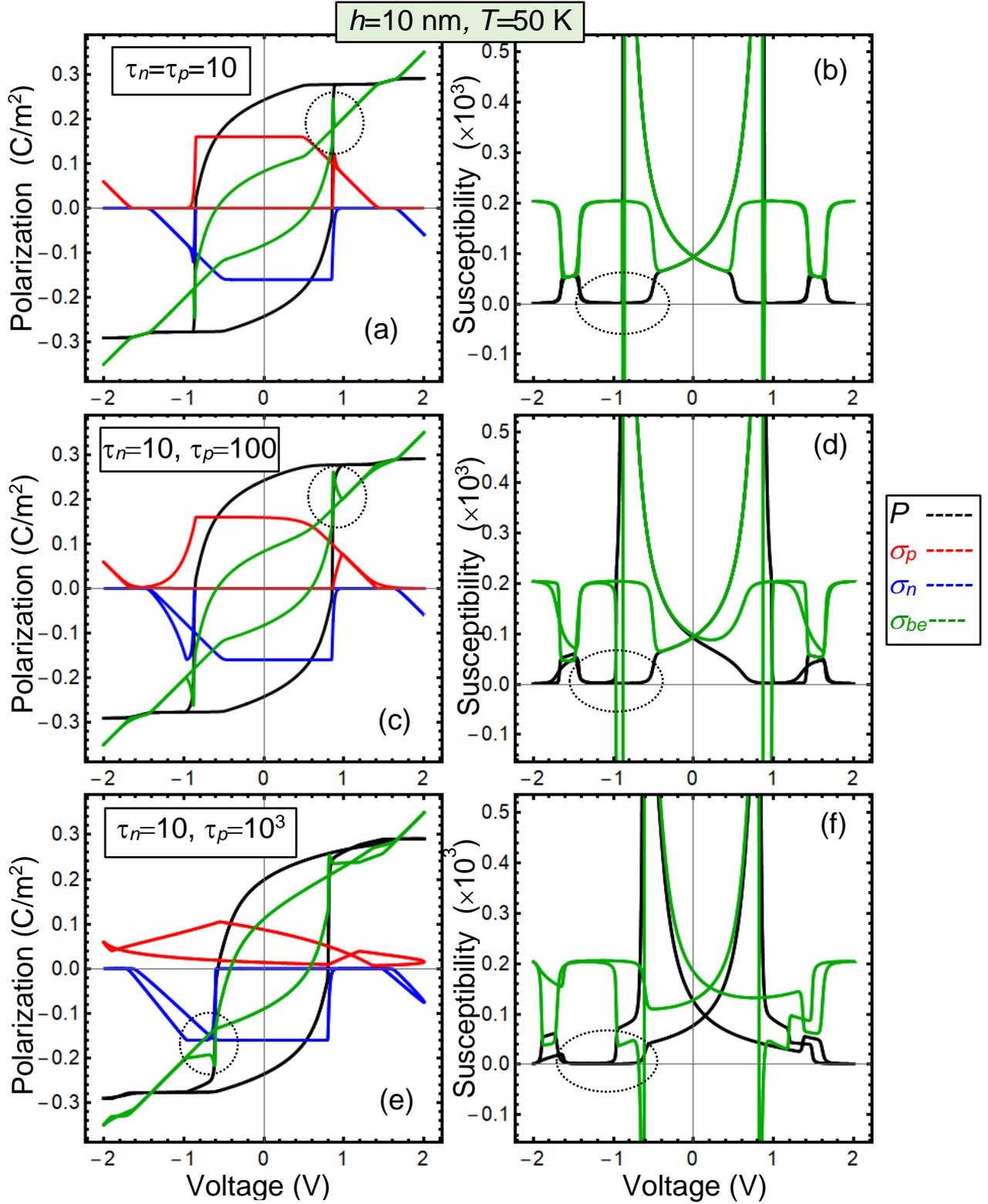

**Figure 9**. **(a, c, e)** The quasi-static dependencies of the polarization (black curves), positive (red curves) and negatives (blue curves) surface charges, and the total charge at bottom electrode (green curves). **(b, d, f)**

The quasi-static dependencies of the dielectric susceptibility (black curves) and effective capacitance (green curves). The dependences are calculated for several values of characteristic times $\tau_n = \tau_p = 10$ **(a, b)**, $\tau_n = 10, \tau_p = 100$ **(c, d)**, and $\tau_n = 10, \tau_p = 10^3$ **(e, f)** in the units of $\tau_{Kh}$. The SPS film thickness $h = 10$ nm, dielectric gap thickness $d = 1.2$ nm, $T = 50$ K. The dimensionless frequency of the quasi-static voltage is $10^{-3}$.

Since the electrode charge can be measured experimentally, its negative derivative can be interpreted as a steady state NC. The peaks-like features of $\sigma_{be}(V)$ and negative $\frac{\partial \sigma_{be}}{\partial V}$ exist in a wide range of film thickness, temperatures and relaxation times $\tau_{n,p} \gg \tau_{Kh}$, being the most pronounced for $\tau_{n,p} \sim 100$. However, the voltage region of negative $\frac{\partial \sigma_{be}}{\partial V}$ is narrow and located very close to the coercive points, $\frac{\partial \bar{P}}{\partial V} = 0$. In contrast to the narrow voltage range, the dielectric susceptibility $\frac{\partial \bar{P}}{\partial V}$ becomes almost zero in a wide voltage range, especially in thin SPS films [see zero parts of black curves in **Figs. 8(b), (d), (f)** and **9(b), (d), (f)**].

Note that both peak-like features of $\sigma_{be}(V)$ and negative $\frac{\partial \sigma_{be}}{\partial V}$, disappear for $\tau_{n,p} \leq 10\tau_{Kh}$, indicating on the sluggish dynamical origin of the apparent NC effect, so it is in fact a very sluggish *transient* NC, but not the "true" steady-state effect. Disregarding the remark one can consider a very-very slow surface charges making the life-time of the transient NC much higher than the period of external bias.

To illustrate the static effects, it makes sense analyze the phase portrait of the free energy, as well as the equilibrium static dependences of polarization and screening charges on applied voltage. The free energy $f_H$, which variation over polarization $\bar{P}$, overpotential $\Psi$ and screening charges $\sigma_{p0} + \sigma_{n0}$ gives the static limit of Eqs.(13), is a sum of the Landau energy $f_P$ and the electrostatic energy $f_E$:

$$f_H[\bar{P}, \Psi, \sigma] = h f_P + f_E, \tag{14a}$$

$$f_P[\bar{P}, \Psi] = \alpha_T (T - T_c) \frac{\bar{P}^2}{2} + \beta \frac{\bar{P}^4}{4} + \gamma \frac{\bar{P}^6}{6} - \frac{\Psi}{h} \bar{P}, \tag{14b}$$

$$f_E[\Psi, \sigma] = -\varepsilon_0 \varepsilon_{33}^b \frac{\Psi^2}{2h} - \varepsilon_0 \varepsilon_d \frac{(\Psi - V)^2}{2d} + \int_0^\Psi (\sigma_{p0}[\varphi] + \sigma_{n0}[\varphi]) d\varphi. \tag{14c}$$

The dependences of the free energy $f_H$ on polarization $\bar{P}$ calculated for several voltages is shown in **Fig. 10(a)** for a thin SPS film and low temperature, and in **Fig. 10(b)** for a thick SPS film and room temperature. Black curves represent the free energy profile at zero voltage calculated with (solid curves) and without (dashed curves) surface screening charges. Since the dashed curves have parabolic shape, while the solid curves have two-well shape at $V = 0$, the screening charges stabilizes the two-well potential relief $f_H(\bar{P})$ of the free energy providing a strong support for the ferroelectric state conservation in thin and ultra-thin films. Applied voltage makes of one well deeper and another shallower, but the depth of the deeper well can be almost voltage-independent [see red and blue curves in **Fig. 10(a)** and **10(b)**]. Free energy dependences on the polarization and applied voltage, which have very flat central saddles, shown by contour lines in **Fig. 10(e)** and **10(f)**, contain multiple almost straight contour lines, which could lead to the voltage-independent parts of $\bar{P}(V)$ curves.

The voltage dependences of polarization $\bar{P}(V)$, screening charges $\sigma_n(V)$, $\sigma_p(V)$ and electrode charge $\sigma_{be}(V)$ are shown in **Fig. 10(c)** and **(d).** Dashed black curves are hysteresis less $\bar{P}(V)$ in the absence of the free charges. Solid black curves show that the screening supports two pronounced bistable polar states at the $\bar{P}(V)$ curves. The voltage-independent regions of the $\bar{P}(V)$ curves are responsible for the appearance of zero derivative $\frac{\partial \bar{P}}{\partial V}$. The unusual voltage dependence of the electrode charge $\sigma_{be}(V)$ with four (for $h = 10$ nm) or two (for $h = 100$ nm) turning points located near the coercive and zero voltages [see green curves in **Fig. 9(a)** and **9(b)**] can explain the negative derivative $\frac{\partial \sigma_{be}}{\partial V}$ shown in **Figs. 8** and **9**.

Hence as follows from **Figs. 8-10**, the addition of the sluggish nonlinear screening charge at the ferroelectric-dielectric interface could be one of the mechanisms to induce a transient NC in the heterostructure.

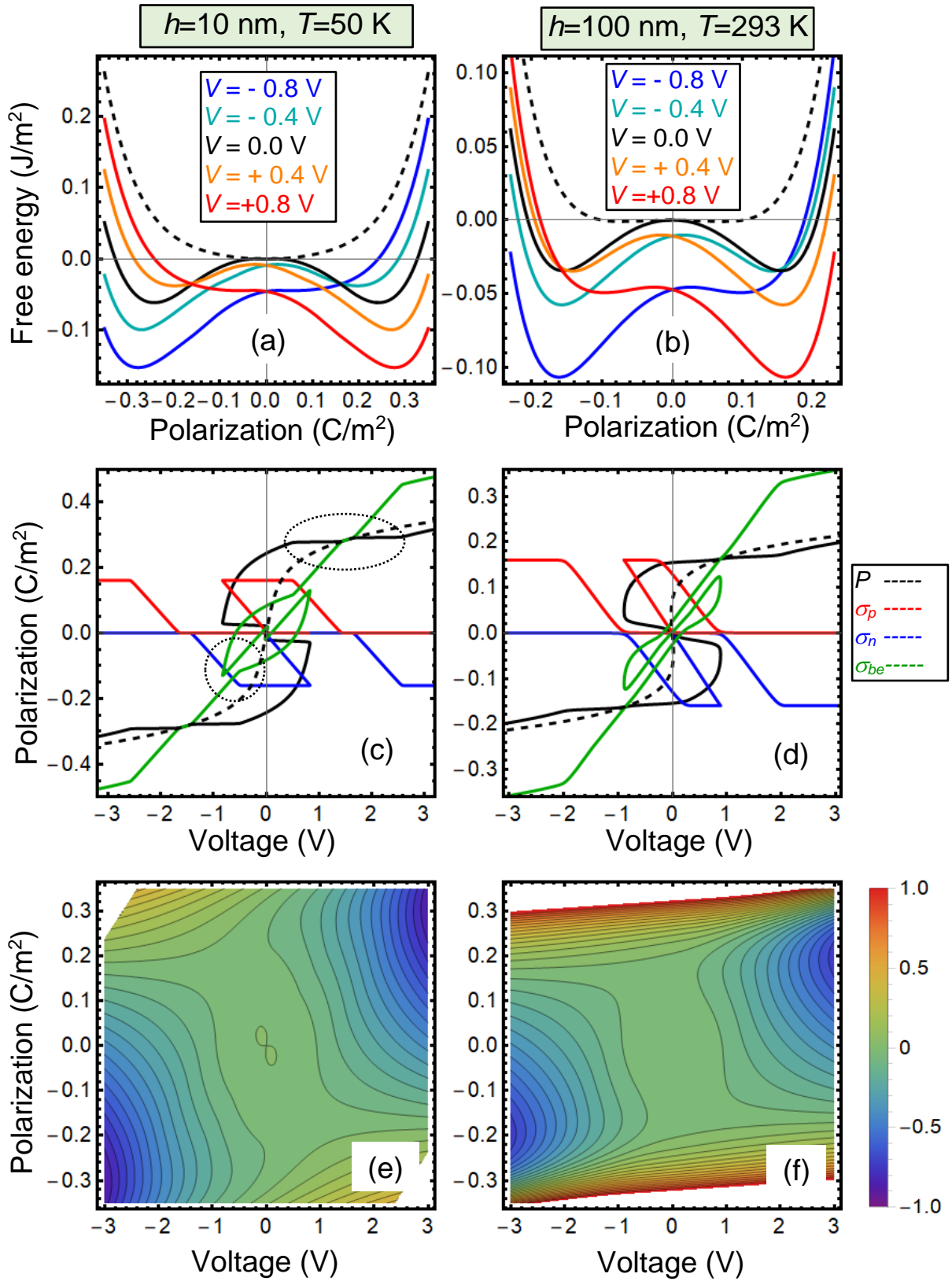

**Figure 10**. **(a, b)** The dependencies of free energy on the polarization for the different values of applied voltage, -0.8, -0.4, 0, 0.4 and 0.8 V (blue, cyan, black, orange and red solid curves, respectively). Dashed black curves represent the free energy profile in the absence of free charges and zero voltage. **(c, d)** The static dependencies of the polarization (black curves), positive (red curves) and negative (blue curves)

surface charges, and total charge at the bottom electrode (green curves). Dashed black curves represent the polarization voltage dependence in the absence of screening charges $\sigma$. **(e, f)** Free energy dependence on the polarization and applied voltage. The dielectric gap thickness $d$=1.2 nm; SPS film thickness $h$ =10 nm and temperature $T = 50$ K for plots **(a, c, e)**; $h$ =100 nm and $T = 293$ K for plots **(b, d, f)**.

## V. Conclusions

Using Landau-Ginzburg-Devonshire approach, we calculate the steady-state and transient capacitance in a uniaxial ferroelectric film covered with a dielectric layer. We consider three basic states of the film: a single-domain, a poly-domain states, and a "ferro-ionic" state, when a single-domain film is covered by a layer of sluggish screening charge, which density of states is a nonlinear function of electric potential.

We derived analytical expressions for the steady-state capacitance of a single-domain state, and study the state stability vs. the domain splitting, which appears under the dielectric layer thickness increase. Analytical expressions for the critical thickness of the dielectric layer, dielectric susceptibility, polarization amplitude and equilibrium domain period at the domain formation onset are also derived. They are very accurate for a single-domain state and at the domain formation onset in a uniaxial ferroelectric film, and illustrated by finite element modelling for thin $Sn_2P_2S_6$ films. The ferroelectric multiaxiality can principally change the domain morphology due to the polarization rotation making the equilibrium domain configuration much more complex and versatile (e.g., vortex-type or flux-closure) and become a possible a reason of NC effect.

The addition of the nonlinear screening charge layer at the ferroelectric-dielectric interface under certain conditions significantly complicates the situation. It appears that when positive or negative, or both kinds of these screening charges are very sluggish in comparison with polarization response to external electric field, the strong retardation between the bound and the screening charge occurs during the voltage cycling. The retardation leads to a very specific peak-like features on the electrode charge hysteresis loops, which derivative with respect to the applied voltage is negative near the coercive voltage. Since the charge is measured experimentally, its negative derivative can be interpreted as a transient state NC. The NC effect is accompanied by almost zero dielectric susceptibility in a wide voltage range. Obtained results may help to elucidate the observability of the transient NC in thin ferroelectric films.

**Acknowledgements.** Authors are grateful to Dr. Pavlo Zubko for useful discussion. A.N.M work is supported by the National Academy of Sciences of Ukraine. This effort is based upon work supported by the U.S. Department of Energy, Office of Science, Office of Basic Energy Sciences Energy Frontier Research Centers program under Award Number DE-SC0021118 (S.V.K.).

**Data availability statement.** The calculations were performed and visualized in Mathematica 12.2 ®Wolfram Research software (https://notebookarchive.org/2021-12-cwwdzxu).

**Supplementary Materials** contains Appendices A, B, C and D with calculation details

**Authors' contribution.** E.A.E., A.N.M. and S.V.K. generated the research idea. A.N.M. formulated the problem mathematically. M.E.Y. derived analytical expressions and obtained numerical results presented in Section II, E.A.E. derived the analytical expressions and performed FEM, which are presented in Section III. A.N.M. derived analytical expressions and generated numerical results using E.A.E. codes, which are presented in Section IV. All co-authors worked on the results discussion, manuscript writing and improvement.

**Supplementary Materials**

**Appendix A. The basic expression derivation for the heterosystem capacitance**

Let us consider a thin ferroelectric film coupled with a passive/dielectric/dielectric layer, with a "background" dielectric permittivity $\varepsilon_b$ and thickness $h$, while dielectric layer has a permittivity $\varepsilon_e$ and thickness $d$ (see **Fig. 2(a)** in the main text). The ferroelectric is characterized by an inhomogeneous spontaneous polarization $\vec{P} = (0, 0, P_3(z))$ perpendicular to the film surfaces. The free energy of the system has the following form

$$G = \int_{z\in[0,h]} \left(\frac{\alpha}{2}P_3^2 + \frac{\beta}{4}P_3^4 + \frac{\gamma}{6}P_3^6 - P_m E_m^{(i)} - \frac{\varepsilon_0 \varepsilon_b}{2} E_m^{(i)} E_m^{(i)} + \frac{g_{11}}{2}\left(\frac{\partial P_3}{\partial z}\right)^2 + \frac{g_{44}}{2}\left[\left(\frac{\partial P_3}{\partial y}\right)^2 + \left(\frac{\partial P_3}{\partial x}\right)^2\right]\right) d^3r + \int_{z\in[-d,0]} d^3r \left(-\frac{\varepsilon_0 \varepsilon_e}{2} E_m^{(e)} E_m^{(e)}\right) \quad (A.1)$$

Here the first term is the energy of ferroelectric, including the electrostatic field and polarization gradient contributions, the second term is the electrostatic energy of the dielectric layer. The gradient coefficients $g_{11}$ and $g_{44}$ determine the correlation radius.

The displacement field is given by the following expressions:

$$\vec{D}^{(i)} = \varepsilon_0 \varepsilon_b \vec{E}^{(i)} + \vec{P}, \quad (A.2)$$

$$\vec{D}^{(e)} = \varepsilon_0 \varepsilon_e \vec{E}^{(e)}, \quad (A.3)$$

inside the ferroelectric film and dielectric layer respectively. Here $\varepsilon_0$ is the universal electric constant; $\vec{E}^{(i)}$ and $\vec{E}^{(e)}$ are the fields inside the film and the dielectric layer, respectively. The minimization of the free energy with respect to polarization gives the equation of state for the polarization and potential, namely

$$\alpha P_3 + \beta P_3^3 + \gamma P_3^5 - g_{11}\frac{\partial^2 P_3}{\partial z^2} - g_{44}\frac{\partial^2 P_3}{\partial y^2} - g_{44}\frac{\partial^2 P_3}{\partial x^2} = E_3^{(i)}, \quad (A.4)$$

The divergence of the displacement equals to zero:

$$\text{div }\vec{D} = 0, \quad (A.5)$$

Below we introduce the electrostatic potential $\varphi$ determining the electric field, $\vec{E} = -\nabla\varphi$. In supposition that the distributions of the polarization and the potential depend only on the $z$-coordinate, we can obtain from Eqs. (A.2) - (A.5) the following equations for potentials:

$$\varepsilon_0 \varepsilon_b \frac{\partial^2 \varphi_i(z)}{\partial z^2} = \frac{\partial P_3(z)}{\partial z}, \quad (A.6)$$

$$\varepsilon_0 \varepsilon_e \frac{\partial^2 \varphi_e(z)}{\partial z^2} = 0. \quad (A.7)$$

Corresponding boundary conditions can be written as follows:

$$\varphi_e(0) - \varphi_i(0) = 0, \quad (A.8)$$

$$D_n^{(e)}(0) - D_n^{(i)}(0) = \sigma, \tag{A.9}$$

$$\varphi_e(-d) = V, \varphi_i(h) = 0. \tag{A.10}$$

The sum of the partial and general solutions of (A.6)-(A.7) is shown below:

$$\varphi_i(z) = -\frac{1}{\varepsilon_0 \varepsilon_b} \int_z^h P_3(\tilde{z}) d\tilde{z} - \frac{h-z}{\varepsilon_0 \varepsilon_b} C_1, \tag{A.11}$$

$$\varphi_e(z) = \frac{C_2 z}{\varepsilon_0 \varepsilon_e} + \frac{C_3}{\varepsilon_0 \varepsilon_e}, \tag{A.12}$$

where $C_1$, $C_2$, $C_3$ are the integration constants to be found from the solution of the system (A.8)-(A.10). In this way the potentials (A.11) and (A.12) can be written as is shown below:

$$\varphi_i(z) = \frac{\bar{P}}{\varepsilon_0 \varepsilon_b} \cdot h \frac{(h-z)\varepsilon_e}{h\varepsilon_e + d\varepsilon_b} - \frac{1}{\varepsilon_0 \varepsilon_b} \int_z^h P_3(\tilde{z}) d\tilde{z} - \frac{(z-h)\varepsilon_e}{\varepsilon_e h + \varepsilon_b d} V + \frac{h-z}{\varepsilon_0 \varepsilon_b} \frac{\varepsilon_b d}{\varepsilon_e h + \varepsilon_b d} \sigma, \tag{A.13}$$

$$\varphi_e(z) = (z+d)\left(-\frac{\bar{P}}{\varepsilon_e \varepsilon_0} \cdot \frac{h \varepsilon_e}{h\varepsilon_e + d\varepsilon_b} - \frac{V \varepsilon_b}{h\varepsilon_e + d\varepsilon_b} + \frac{\varepsilon_e h}{\varepsilon_e h + \varepsilon_b d} \frac{\sigma}{\varepsilon_0 \varepsilon_e}\right) + V, \tag{A.14}$$

where $\bar{P} = \frac{1}{h}\int_0^h P_3(\tilde{z}) d\tilde{z}$ is the mean value of the film polarization. Displacement components are $D_i = \frac{\varepsilon_b d}{\varepsilon_e h + \varepsilon_b d} P_3 + \frac{\varepsilon_b d}{\varepsilon_e h + \varepsilon_b d} \sigma$ and $D_e = \frac{\varepsilon_e h}{h\varepsilon_e + d\varepsilon_b}(P_3 - \sigma)$. The charge at the bottom electrode is $\sigma_{be} = \frac{\varepsilon_e h}{\varepsilon_e h + \varepsilon_b d}(P_3 - \sigma)$, and the charge at the top electrode is $\sigma_{te} = -\frac{\varepsilon_e h}{\varepsilon_e h + \varepsilon_b d} P_3 - \frac{\varepsilon_b d}{\varepsilon_e h + \varepsilon_b d} \sigma$. The total charge is zero $\sigma_{be} + \sigma_{te} + \sigma = 0$.

For the purposes of this **Appendix**, we further put $\sigma = 0$. The distribution of the electric field inside dielectric and ferroelectric layers can be obtained from Eqs. (A.13) and (A.14):

$$E^{(i)}(z) = -\frac{P(z)}{\varepsilon_0 \varepsilon_b} + \frac{\bar{P}}{\varepsilon_0 \varepsilon_b} \cdot \frac{h\varepsilon_e}{h\varepsilon_e + d\varepsilon_b} + \frac{V\varepsilon_e}{h\varepsilon_e + d\varepsilon_b}, \tag{A.15}$$

$$E^{(e)}(z) = \frac{\bar{P}}{\varepsilon_e \varepsilon_0} \cdot \frac{h\varepsilon_e}{h\varepsilon_e + d\varepsilon_b} + \frac{V\varepsilon_b}{h\varepsilon_e + d\varepsilon_b}. \tag{A.16}$$

Below we suppose that the ferroelectric polarization is constant inside the film $P(z) \equiv \bar{P}$. Neglecting the gradient terms and substituting the field (A.15) in the equation (A.4) for the polarization gives the following:

$$\left(\alpha + \frac{d}{\varepsilon_0(d\varepsilon_b + h\varepsilon_e)}\right)\bar{P} + \beta\bar{P}^3 + \gamma\bar{P}^5 = \frac{\varepsilon_e V}{d\varepsilon_b + h\varepsilon_e} \tag{A.17}$$

Free energy for the case of a homogeneous polarization is shown below

$$\frac{G}{S} = h\left[\left(\frac{\alpha}{2} + \frac{d}{2\varepsilon_0(d\varepsilon_b + h\varepsilon_e)}\right)\bar{P}^2 + \frac{\beta}{4}\bar{P}^4 + \frac{\gamma}{6}\bar{P}^6\right] - \frac{h\varepsilon_e}{d\varepsilon_b + h\varepsilon_e} V\bar{P} - \frac{\varepsilon_0 \varepsilon_b \varepsilon_e V^2}{2(d\varepsilon_b + h\varepsilon_e)} \tag{A.18}$$

The equation (A.17) is an algebraic equation of the fifth order; hence, its solution could not be found in the closed form. However, the mean polarization at zero bias can be written as follows:

$$\bar{P} = \pm\sqrt{\frac{1}{2\gamma}\left[-\beta \pm \sqrt{\beta^2 - 4\gamma\left(\alpha + \frac{d}{\varepsilon_0(d\varepsilon_b + h\varepsilon_e)}\right)}\right]} \tag{A.19}$$

Differentiation of both sides of (A.17) with respect to the bias gives the following:

$$\frac{d\bar{P}}{dV} = \frac{\varepsilon_0 \varepsilon_e}{\varepsilon_0(d\varepsilon_b + h\varepsilon_e)(\alpha + 3\beta\bar{P}^2 + 5\gamma\bar{P}^4) + d} \quad (A.20)$$

Let us find the surface charge density on the electrodes allowing one to find the capacitance using the formula $C_{surf} = S \frac{d\sigma_{el}}{dV}$. It is known that the displacement field near metal electrode determines the surface charge on it:

$$D_n^{(e)}(-d) = \sigma_{el} \quad \text{and} \quad D_n^{(i)}(-h) = -\sigma_{el} \quad (A.21)$$

Using Eqs.(A.2), (A.4) as well as (A.16), the surface charge can be found as follows:

$$\sigma_{el} = \bar{P} \cdot \frac{h\varepsilon_e}{h\varepsilon_e + d\varepsilon_b} + \frac{\varepsilon_b \varepsilon_e \varepsilon_0 V}{h\varepsilon_e + d\varepsilon_b}. \quad (A.22)$$

Thus, using (A.22), the capacitance per unit area can be written as is shown below:

$$\frac{C_{surf}}{S} = \frac{\varepsilon_0 \varepsilon_e}{\varepsilon_0(\varepsilon_e h + \varepsilon_b d)(\alpha + 3\beta\bar{P}^2 + 5\gamma\bar{P}^4) + d} \cdot \frac{h\varepsilon_e}{h\varepsilon_e + d\varepsilon_b} + \frac{\varepsilon_b \varepsilon_e \varepsilon_0}{h\varepsilon_e + d\varepsilon_b} \quad (A.23)$$

One could interpret Eq.(A.23) in a way that the later term is the well-known capacitance of a flat capacitor filled with "simple" dielectrics, characterized by their permittivity values $\varepsilon_b$ and $\varepsilon_e$, while the former one is directly related to ferroelectric contribution, renormalized under the size effect influence. Let us perform some transformations of Eq.(A.23):

$$\frac{C_{surf}}{S} = \left[ \frac{\frac{h}{\varepsilon_b}}{\varepsilon_0 \varepsilon_b (\alpha + 3\beta\bar{P}^2 + 5\gamma\bar{P}^4)\left(\frac{d}{\varepsilon_e} + \frac{h}{\varepsilon_b}\right) + \frac{d}{\varepsilon_e}} + 1 \right] \frac{\varepsilon_0}{\frac{h}{\varepsilon_b} + \frac{d}{\varepsilon_e}} = \frac{\varepsilon_0}{\frac{d}{\varepsilon_e} + \frac{h}{\varepsilon_{eff}}}, \quad (A.24a)$$

$$\varepsilon_{eff} = \varepsilon_b + \frac{1}{\varepsilon_0 \{\alpha + 3\beta\bar{P}^2 + 5\gamma\bar{P}^4\}} \quad (A.24b)$$

However, despite the superficial analogy with the series capacitance formula, it should be noted that the expression $\varepsilon_b + \frac{1}{\varepsilon_0(\alpha + 3\beta\bar{P}^2 + 5\gamma\bar{P}^4)}$ has nothing to do with permittivity of ferroelectric film, coupled to dielectric layer (compare with (A.20)). According to the stability condition, obtained from Eq. (A.17) and (A.18)

$$\alpha + 3\beta\bar{P}^2 + 5\gamma\bar{P}^4 > -\frac{d}{\varepsilon_0(d\varepsilon_b + h\varepsilon_e)}. \quad (A.25)$$

the expression $\varepsilon_b + \frac{1}{\varepsilon_0(\alpha + 3\beta\bar{P}^2 + 5\gamma\bar{P}^4)}$ could even have a negative sign, provoking the thoughts about "negative" capacitance of a ferroelectric film.

**Appendix B. The derivation of the temperature of the break into domain structure**

**B.1. Problem statement**

The linearized system of equations for polarization and electric potential inside the ferroelectric film and outside it has the following form (see (A.4) and (A.5)):

$$\alpha P_3 - g_{11}\frac{\partial^2 P_3}{\partial z^2} - g_{44}\left(\frac{\partial^2 P_3}{\partial x^2} + \frac{\partial^2 P_3}{\partial y^2}\right) = -\frac{\partial \phi}{\partial z}, \tag{B.1a}$$

$$\left(\frac{\partial^2}{\partial x^2} + \frac{\partial^2}{\partial y^2} + \frac{\partial^2}{\partial z^2}\right)\phi^{(in)} = \frac{1}{\varepsilon_0 \varepsilon_b}\frac{\partial P_3}{\partial z}, \tag{B.1b}$$

$$\left(\frac{\partial^2}{\partial x^2} + \frac{\partial^2}{\partial y^2} + \frac{\partial^2}{\partial z^2}\right)\phi^{(out)} = 0. \tag{B.1c}$$

Let us consider harmonic-like fluctuations

$$P_3 = P_k(z)\exp(i\,k\,x), \quad \phi^{(in)} = \phi_k^{(in)}(z)\exp(i\,k\,x), \quad \phi^{(out)} = \phi_k^{(out)}(z)\exp(ikx). \tag{B.2}$$

The equations for the amplitudes could be easily found from Eqs.(B.1):

$$(\alpha + g_{44}k^2)P_k - g_{11}\frac{\partial^2 P_k}{\partial z^2} = -\frac{\partial \phi_k}{\partial z}, \tag{B.3a}$$

$$\frac{\partial^2 \phi_k^{(in)}}{\partial z^2} - k^2 \phi_k^{(in)} = \frac{1}{\varepsilon_0 \varepsilon_b}\frac{\partial P_k}{\partial z}, \tag{B.3b}$$

$$\frac{\partial^2 \phi_k^{(out)}}{\partial z^2} - k^2 \phi_k^{(out)} = 0. \tag{B.3c}$$

Differentiation of the Eqs. (B.3) gives the following expression

$$\left(\frac{\partial^2}{\partial z^2} - k^2\right)\left[(\alpha + g_{44}k^2)P_k - g_{11}\frac{\partial^2 P_k}{\partial z^2}\right] = \left(\frac{\partial^2}{\partial z^2} - k^2\right)\left[-\frac{\partial \phi_k}{\partial z}\right], \tag{B.4a}$$

$$\frac{\partial}{\partial z}\left(\frac{\partial^2}{\partial z^2} - k^2\right)\phi_k^{(in)} = \frac{1}{\varepsilon_0 \varepsilon_b}\frac{\partial^2 P_k}{\partial z^2}. \tag{B.4b}$$

Hence, one could exclude the potential amplitude from Eq. (B.4a) and get the equation for the polarization amplitude in the following form:

$$\left(\frac{\partial^2}{\partial z^2} - k^2\right)\left[(\alpha + g_{44}k^2)P_k - g_{11}\frac{\partial^2 P_k}{\partial z^2}\right] = -\frac{1}{\varepsilon_0 \varepsilon_b}\frac{\partial^2 P_k}{\partial z^2} \tag{B.5}$$

Let us look for the solution of (B.5) in the form of $P_3 \sim exp(qz)$, where the inverse characteristic length $q$ satisfies the following equation:

$$(q^2 - k^2)(\alpha + g_{44}k^2 - g_{11}q^2) = -\frac{q^2}{\varepsilon_0 \varepsilon_b} \tag{B.6a}$$

$$q^4 - \left(\frac{\alpha + g_{44}k^2}{g_{11}} + k^2 + \frac{1}{\varepsilon_0 \varepsilon_b g_{11}}\right)q^2 + \frac{\alpha + g_{44}k^2}{g_{11}}k^2 = 0 \tag{B.6b}$$

Its solutions could be written as

$$q_{1,2}^2 = \frac{1}{2}\left( \frac{\alpha+g_{44}k^2}{g_{11}} + k^2 + \frac{1}{\varepsilon_0\varepsilon_b g_{11}} \pm \sqrt{\left(\frac{\alpha+g_{44}k^2}{g_{11}} + k^2 + \frac{1}{\varepsilon_0\varepsilon_b g_{11}}\right)^2 - 4\frac{\alpha+g_{44}k^2}{g_{11}}k^2} \right) \quad \text{(B.6c)}$$

It should be noted that in most cases $\varepsilon_0\varepsilon_b g_{11} \ll \{1/k^2, g_{11}/|\alpha|, g_{44}/|\alpha|\}$, hence the following approximations are valid

$$q_1 \approx k\sqrt{\frac{\alpha+g_{44}k^2}{\alpha+g_{44}k^2+g_{11}k^2+\frac{1}{\varepsilon_0\varepsilon_b}}}, \quad q_2 \approx \sqrt{\frac{1}{\varepsilon_0\varepsilon_b g_{11}} - q_1^2} \quad \text{(B.6d)}$$

### B.2. The system with a dielectric layer of finite thickness "d"

In this case the boundary and interface conditions have the following form:

$$\left(\frac{\partial P_3}{\partial z}\right)\bigg|_{z=0,h} = 0, \quad \text{(B.7a)}$$

$$\left(\phi^{(out)}\right)\bigg|_{z=-d} = 0 \quad \text{(B.7b)}$$

$$\left(-\varepsilon_0\varepsilon_b \frac{\partial \phi^{(in)}}{\partial z} + P_3 + \varepsilon_0\varepsilon_e \frac{\partial \phi^{(out)}}{\partial z}\right)\bigg|_{z=0} = 0. \quad \text{(B.7c)}$$

$$\left(\phi^{(out)} - \phi^{(in)}\right)\bigg|_{z=0} = 0 \quad \text{(B.7d)}$$

$$\left(\phi^{(in)}\right)\bigg|_{z=h} = 0 \quad \text{(B.7e)}$$

Now we could write the general solution of Eq.(B.5) as is shown below:

$$P_k = s_1 \sinh(q_1 z) + s_2 \sinh(q_2 z) + c_1 \cosh(q_1 z) + c_2 \cosh(q_2 z) \quad \text{(B.8a)}$$

The corresponding solution for the potential is

$$\phi_k^{(in)} = g_1 \cosh(q_1 z) + g_2 \cosh(q_2 z) + f_1 \sinh(q_1 z) + f_2 \sinh(q_2 z) \quad \text{(B.8b)}$$

and

$$\phi_k^{(out)} = g\frac{\sinh(k(z+d))}{\sinh(k\,d)} \quad \text{(B.8c)}$$

This function already satisfies the condition (B.7b).

Using Eq.(B.2b) and the independence of different solutions from (B.8a), one could easily find the following relations

$$c_i = \varepsilon_0\varepsilon_b \frac{q_i^2-k^2}{q_i} f_i, \quad s_i = \varepsilon_0\varepsilon_b \frac{q_i^2-k^2}{q_i} g_i \quad \text{(B.9)}$$

The four constants $s_i$ and $c_i$ should be found from boundary conditions (B.2).

The conditions (B.7a) $((\partial P_k/\partial z)|_{z=0,h} = 0)$ give the following equations:

$$q_1 s_1 + q_2 s_2 = 0 \tag{B.10a}$$

$$q_1 s_1 \cosh(q_1 h) + q_2 s_2 \cosh(q_2 h) + q_1 c_1 \sinh(q_1 h) + q_2 c_2 \sinh(q_2 h) = 0 \tag{B.10b}$$

And the condition (B.7e), $\left(\phi^{(in)}\right)\big|_{z=h} = 0$ yields the following:

$$g_1 \cosh(q_1 h) + g_2 \cosh(q_2 h) + f_1 \sinh(q_1 h) + f_2 \sinh(q_2 h) = 0 \tag{B.10c}$$

In turn, the condition (B.7d) $\left(\phi^{(out)} - \phi^{(in)}\right)\big|_{z=0} = 0$ leads to the following

$$g = g_1 + g_2 \tag{B.10d}$$

Finally, the condition (B.7c) $\left(-\varepsilon_0 \varepsilon_b \frac{\partial \phi^{(in)}}{\partial z} + P_3 + \varepsilon_0 \varepsilon_e \frac{\partial \phi^{(out)}}{\partial z}\right)\big|_{z=0} = 0$ transforms into

$$-\varepsilon_0 \varepsilon_b (q_1 f_1 + q_2 f_2) + c_1 + c_2 + |k|\varepsilon_0 \varepsilon_e \frac{\cosh(|k|d)}{\sinh(|k|d)} g = 0. \tag{B.10e}$$

Eqs.(B.10) supplemented with (B.9) presents the linear system for the unknown coefficients. It could be easily reduced to the system of four linear equations for the $s_i$ and $c_i$ constants. The formal solution is zero, but since we have a homogeneous system, we are interested in the stability analysis. Hence, we should look for a zero point of the corresponding linear equations system determinant for $s_i$ and $c_i$.

$$\frac{q_1 q_2 (q_1^2 - q_2^2)}{(k^2 - q_1^2)(k^2 - q_2^2)} \left[ \frac{k^2 q_2}{k^2 - q_1^2} \cosh(q_1 h) \sinh(q_2 h) - \frac{k^2 q_1}{k^2 - q_2^2} \cosh(q_2 h) \sinh(q_1 h) + \right.$$

$$\left. \frac{\varepsilon_e}{\varepsilon_b} \frac{q_1 q_2 (q_1^2 - q_2^2) k \coth(d\,k)}{(k^2 - q_1^2)(k^2 - q_2^2)} \sinh(q_1 h) \sinh(q_2 h) \right] = 0 \tag{B.11a}$$

It is easy to show that

$$q_2 \gg q_1 \text{ and } q_2 h \gg 1 \text{ (see (B.6c))}$$

for the most of the reasonable cases. Hence, the first multiplier could be dropped, while the rest of Eq.(B.11a) could be simplified using the common relation $\sinh(q_2 h) \approx \cosh(q_2 h)$

$$\frac{k^2 q_2}{k^2 - q_1^2} \cosh(q_1 h) - \frac{k^2 q_1}{k^2 - q_2^2} \sinh(q_1 h) + \frac{\varepsilon_e}{\varepsilon_b} \frac{q_1 q_2 (q_1^2 - q_2^2) k \coth(d\,k)}{(k^2 - q_1^2)(k^2 - q_2^2)} \sinh(q_1 h) = 0$$

$$\tag{B.11b}$$

$$\frac{k^2 - q_2^2}{q_2^2} \cosh(q_1 h) + \frac{(k^2 - q_1^2)}{q_2^3} q_1 \sinh(q_1 h) + \frac{\varepsilon_e}{\varepsilon_b} \frac{q_1 (q_2^2 - q_1^2) k \coth(d\,k)}{q_2^2 k^2} \sinh(q_1 h) = 0$$

Next, we appeal to the strong inequalities $q_2 \gg q_1$ and $q_2 \gg k$ and use the evident form of $q_1$ and $q_2$:

$$(1 - g_{11}k^2\varepsilon_0\varepsilon_b)\cosh(q_1 h) + \frac{g_{11}k^2}{\alpha + g_{44}k^2 + \frac{1}{\varepsilon_0\varepsilon_b}}\sqrt{\varepsilon_0\varepsilon_b g_{11}}\, q_1 \sinh(q_1 h)$$

$$+ \left(1 - g_{11}k^2\varepsilon_0\varepsilon_b + \frac{g_{11}k^2}{\alpha + g_{44}k^2 + \frac{1}{\varepsilon_0\varepsilon_b}}\right)\frac{\varepsilon_e}{\varepsilon_b}\frac{q_1 k \coth(d\,k)}{k^2}\sinh(q_1 h) = 0$$

(B.11c)

Next, we dropped the second term and used that $g_{11}k^2\varepsilon_0\varepsilon_b \ll 1$:

$$\cosh(q_1 h) + \frac{\varepsilon_e}{\varepsilon_b}\frac{\coth(kd)}{k}q_1 \sinh(q_1 h) = 0 \qquad (B.11d)$$

Since all the terms in Eq.(B.11d) are formally positive, it could be satisfied only for an imaginary $q_1 \equiv i\,q$, therefore $\cosh(q_1 h) = \cos(qh)$ and $q_1 \sinh(q_1 h) = -q\sin(qh)$:

$$\cos(qh) - \frac{\varepsilon_e q}{\varepsilon_b k}\coth(kd)\sin(qh) = 0 \qquad (B.11e)$$

The equation (B.11e) represents the boundary of the stability region of the paraelectric phase with respect to the appearance of the domain structure. Taking into account linear temperature dependence of α, one could consider this equation as the dependence of the transition temperature on the thickness of ferroelectric film, dielectric layer and domain structure period (wave vector $k$). It is usually assumed that the transition temperature should be maximized with respect to the latter (in reality the transition takes place at the highest possible transition temperature). Analytical calculations are not possible for the general form of Eq.(B.11e), therefore below we consider several limiting cases.

**Case I. k→0**

Expansion of Eq.(B.11e) in series in the limit $q\,h \to 0$ gives

$$1 - \frac{q^2 h^2}{2} - \frac{\varepsilon_e h}{\varepsilon_b k}\coth(kd)\, q^2 = 0 \qquad (B.12a)$$

and finally, using evident form of parameter $q \approx k\sqrt{\frac{-\alpha - g_{44}k^2}{\alpha + g_{44}k^2 + \frac{1}{\varepsilon_0\varepsilon_b}}}$, one could rewrite Eq.(B.12a) in

the form of

$$\alpha + g_{44}k^2 + \frac{1}{\varepsilon_0\varepsilon_b\left[1 + \frac{k^2 h^2}{2} + \frac{\varepsilon_e k h}{\varepsilon_b}\coth(kd)\right]} = 0, \qquad (B.12b)$$

which for the case of a thin dielectric layer could be reduced to the following

$$\alpha + g_{44}k^2 + \frac{1}{\varepsilon_0\varepsilon_b\left[1 + \frac{k^2 h^2}{2} + \frac{\varepsilon_e h}{\varepsilon_b d}\right]} \approx 0. \qquad (B.12c)$$

Minimization of Eq.(B.12c) with respect to the wave vector gives equation

$$\varepsilon_0\varepsilon_b d\alpha + \varepsilon_0\varepsilon_b g_{44} 2kdk - \frac{k h^2 dk}{\left[1+\frac{k^2 h^2}{2}+\frac{\varepsilon_e h}{\varepsilon_b d}\right]^2} \approx 0 \quad \text{(B.12d)}$$

Condition $d\alpha/dk = 0$ gives the domain structure equilibrium wave vector near the phase transition:

$$k_{eq} = \sqrt{\frac{1}{h}\sqrt{\frac{2}{\varepsilon_0\varepsilon_b g_{44}}} - \frac{2}{h^2}\left(1+\frac{\varepsilon_e h}{\varepsilon_b d}\right)} \quad \text{(B.12e)}$$

Let us check the condition $qh \ll 1$, which is ground for the derivation of Eq.(B.12a)

$$qh = \sqrt{\frac{2h\sqrt{\frac{2}{\varepsilon_0\varepsilon_b g_{44}}}-4\left(1+\frac{\varepsilon_e h}{\varepsilon_b d}\right)}{h\sqrt{\frac{2}{\varepsilon_0\varepsilon_b g_{44}}}-2}} \approx \left|\sqrt{\varepsilon_0\varepsilon_b g_{44}} \ll h\right| \approx \sqrt{2}\sqrt{1-\frac{\varepsilon_e\sqrt{2\varepsilon_0\varepsilon_b g_{44}}}{\varepsilon_b d}} \ll 1 \quad \text{(B.12f)}$$

It can be clearly seen that for a rather thick dielectric layer the approximation (B.12a) is not satisfactory, and it is obvious that (B.12f) takes place only for rather thin layer with a thickness of

$$d \gtrsim \frac{\varepsilon_e\sqrt{2\varepsilon_0\varepsilon_b g_{44}}}{\varepsilon_b} \quad \text{(B.12g)}$$

It should be noted that for the case $d < \varepsilon_e\sqrt{2\varepsilon_0 g_{44}/\varepsilon_b}$ the wave vector (B.12e) becomes imaginary and periodic domain structure is not possible.

Finally, let us substitute the solution (B.12e) into (B.12c) to get the equilibrium stability boundary of the paraelectric phase with respect to the appearance of the domain structure:

$$\alpha + g_{44}\left\{\sqrt{\frac{2}{\varepsilon_0\varepsilon_b g_{44}}} - \left(\frac{1}{h}+\frac{\varepsilon_e}{\varepsilon_b d}\right)\right\}\frac{2}{h} \approx 0 \quad \text{(B.12h)}$$

Note, that if the dielectric layer thickness is smaller than $d < \varepsilon_e\sqrt{2\varepsilon_0 g_{44}/\varepsilon_b}$ (which is actually very small value of order of fraction of lattice constants), this solution could not be realized, since $k_{eq}$ becomes imaginary and the domain structure instability could not be realized. The single domain state with k=0 is reached, which gives the following condition for the phase transition

$$\alpha + \frac{d}{\varepsilon_0(d\varepsilon_b + h\varepsilon_e)} = 0 \quad \text{(B.12i)}$$

**Case II.** $qh > 1$

In the case of the thick layer

$$d \gg \frac{\varepsilon_e\sqrt{2\varepsilon_0\varepsilon_b g_{44}}}{\varepsilon_b} \quad \text{(B.13)}$$

the condition (B.12f) is violated and a more sophisticated approximation is needed. Here we recall the expression for the tangent function, knows as the partial fraction expansion, namely

$$\tan(y) \approx \frac{2y}{\frac{\pi^2}{4}-y^2} \quad \text{at} \quad y^2 \lesssim \pi^2/4 \quad \text{(B.14a)}$$

Here we actually used only one term left from the infinite series. Note that at $y \to 0$ this approximation gives only a rough picture of "true tangent", namely the latter is $\tan(y) \approx y$, while the former gives $\tan(y) \approx 8y/\pi^2$ (with error of about 20%).

Using Eq.(B.14a), Eq.(B.11e) could be transformed as follows

$$1 - \frac{\varepsilon_e h \coth(kd)}{\varepsilon_b k} \frac{q^2}{\frac{\pi^2}{8}-\frac{(qh)^2}{2}} = 0 \quad \text{(B.14b)}$$

And, finally using the evident form of parameter $q \approx k\sqrt{\frac{-\alpha-g_{44}k^2}{\alpha+g_{44}k^2+\frac{1}{\varepsilon_0\varepsilon_b}}}$, one could rewrite Eq.(B.14b) in the form of

$$\alpha + g_{44}k^2 + \frac{\pi^2/8}{\varepsilon_0\varepsilon_b\left(\pi^2/8+\frac{h^2}{2}k^2+\frac{\varepsilon_e \coth(kd)}{\varepsilon_b}h\,k\right)} = 0 \quad \text{(B.14c)}$$

Below we consider the case $kd \ll 1$, so that $\coth(kd) \approx 1/(kd)$. Minimization of (B.14c) with respect to k gives the equation

$$\frac{4\sqrt{\varepsilon_0\varepsilon_b g_{44}}}{\pi h} = \frac{1}{\left(\frac{\pi^2}{8}+\frac{h^2}{2}k^2+\frac{\varepsilon_e h}{\varepsilon_b d}\right)}$$

Solving the above equation, we get the equilibrium wave vector of the domain structure near the phase transition

$$k_{eq} \approx \frac{1}{h}\sqrt{\frac{\pi h}{2\sqrt{\varepsilon_0\varepsilon_b g_{44}}}-\frac{\pi^2}{4}-2\frac{\varepsilon_e h}{\varepsilon_b d}} \quad \text{(B.14d)}$$

This solution is physically reasonable when

$$\frac{\pi h}{2\sqrt{\varepsilon_0\varepsilon_b g_{44}}}-\frac{\pi^2}{4}-2\frac{\varepsilon_e h}{\varepsilon_b d} > 0 \Rightarrow d > \frac{4\varepsilon_e\sqrt{\varepsilon_0\varepsilon_b g_{44}}}{\pi\varepsilon_b\left(1-\frac{\pi\sqrt{\varepsilon_0\varepsilon_b g_{44}}}{2\,h}\right)} \quad \text{(B.14e)}$$

Let us check the condition $q\,h \sim 1$, which is ground for the derivation of Eq.(B.14b)

$$hq \approx \frac{\pi}{2}\sqrt{1-\frac{4}{\pi}\frac{\varepsilon_e h}{\varepsilon_b d}\frac{\sqrt{\varepsilon_0\varepsilon_b g_{44}}}{h\left(1-\frac{\pi\sqrt{\varepsilon_0\varepsilon_b g_{44}}}{2\,h}\right)}} \approx \left|\sqrt{\varepsilon_0\varepsilon_b g_{44}} \ll h\right| \approx \frac{\pi}{2}\sqrt{1-\frac{4}{\pi}\frac{\varepsilon_e}{\varepsilon_b}\frac{\sqrt{\varepsilon_0\varepsilon_b g_{44}}}{d}} \quad \text{(B.14f)}$$

It is seen that for the physically reasonable values of dielectric layer thickness (with the characteristic length $\sqrt{\varepsilon_0\varepsilon_b g_{44}}$ having the value of about few Å) the following relation takes place: $hq \lesssim \pi/2$, which justifies the usage of the approximation for tangent function (B.14a). Note that even near the threshold $d \approx \varepsilon_e\sqrt{2\varepsilon_0 g_{44}/\varepsilon_b}$ the value of $hq \approx 0.25$ is still acceptable for (B.14a). Substitution $k = k_{eq}$ from Eq. (B.14d) into Eq.(B14c) at $kd \ll 1$ gives the condition of the transition into multidomain state for the case of the thin dielectric layer:

$$\alpha + g_{44}\left(\frac{\pi}{\sqrt{\varepsilon_0\varepsilon_b g_{44}}h} - \frac{\pi^2}{4h^2} - 2\frac{\varepsilon_e}{\varepsilon_b d\, h}\right) = 0 \qquad (B.14g)$$

The condition $kd \ll 1$ is satisfied when

$$\frac{d}{h}\sqrt{\frac{\pi\, h}{2\sqrt{\varepsilon_0\varepsilon_b g_{44}}} - \frac{\pi^2}{4} - 2\frac{\varepsilon_e h}{\varepsilon_b d}} \ll 1 \qquad \frac{\pi\, d^2}{2h\sqrt{\varepsilon_0\varepsilon_b g_{44}}} - \frac{\pi^2 d^2}{4h^2} - 2\frac{\varepsilon_e d}{\varepsilon_b h} \ll 1$$

It is seen that the first and the second terms in the right-hand side of the last strong inequality could be dropped to give the required condition in the following form

$$d^2 \ll \frac{2}{\pi} h\sqrt{\varepsilon_0\varepsilon_b g_{44}}. \qquad (B.14h)$$

Thus, the case considered above is applicable for the intermediate value of dielectric layer thickness, namely

$$\frac{4\varepsilon_e\sqrt{\varepsilon_0\varepsilon_b g_{44}}}{\pi\varepsilon_b\left(1 - \frac{\pi\sqrt{\varepsilon_0\varepsilon_b g_{44}}}{2\,h}\right)} < d \ll \sqrt{h\sqrt{\varepsilon_0\varepsilon_b g_{44}}}$$

**Case III.** $kd \gg 1$

For the case of a thick "dielectric layer" Eq. (B.14c) takes the following form

$$\frac{\pi^2}{8\varepsilon_0\varepsilon_b\left(\frac{\pi^2}{8} + \frac{(kh)^2}{2} + \frac{\varepsilon_e}{\varepsilon_b}hk\right)} + \alpha + g_{44}k^2 = 0 \qquad (B.15a)$$

This equation could be easily maximized with respect to $k$ for the case of $hk \gg 1$, to give equilibrium wave vector of the domain structure near the phase transition

$$k_{eq} \approx \frac{1}{h}\sqrt{\frac{\pi\, h}{2\sqrt{\varepsilon_0\varepsilon_b g_{44}}} - \frac{\pi^2}{4}} \qquad (B.15b)$$

The condition $hk \gg 1$ leads to $h \gg \sqrt{\varepsilon_0\varepsilon_b g_{44}}$ which is valid in the majority of the cases. At the same time, the condition $kd \gg 1$ leads to the condition $d \gg \sqrt{h\sqrt{\varepsilon_0\varepsilon_b g_{44}}}$ (compare with (B.14h). Substitution $k = k_{max}$ from Eq. (B.15b) into Eq.(B15a) gives the condition of the transition into multidomain state for the case of the thick dielectric layer:

$$\alpha + g_{44}\left(\frac{\pi}{h\sqrt{\varepsilon_0\varepsilon_b g_{44}}} - \frac{\pi^2}{4h^2}\right) = 0 \qquad (B.15c)$$

**B3. Variational consideration of the domain structure in harmonic approximation**

The harmonic periodic solution of linearized system of equations for polarization and electric potential inside the ferroelectric film and outside it is given by Eqs.(B.2), (B.8a), (B.8b) and (B.8c). Considering homogeneous (**Appendix A**) and harmonic (**Appendix B**) solutions, we could write the trial functions in the following form

$$P_3 = P + 2A \cos(k\,x) \left\{ \cosh(q_1(z-h)) - \frac{q_1 \sinh(q_1 h)}{q_2 \sinh(q_2 h)} \cosh(q_2(z-h)) \right\}, \quad \text{(B.16a)}$$

$$\phi^{(in)} = \frac{-P \frac{d}{\varepsilon_e}(h-z)}{\varepsilon_0 \varepsilon_b \frac{h}{\varepsilon_b}+\frac{d}{\varepsilon_e}} + \frac{2A}{\varepsilon_0 \varepsilon_b} \cos(k\,x) \left\{ \frac{q_1}{q_1^2 - k^2} \sinh(q_1(z-h)) - \frac{q_1 \sinh(q_1 h)}{q_2 \sinh(q_2 h)} \frac{q_2}{q_2^2 - k^2} \sinh(q_2(z-h)) \right\}$$

(B.16b)

Here the first terms represent a homogeneous solution with the magnitude "$P$" (average polarization), corresponding to a single domain state. The last terms represent a harmonically modulated solution with amplitude "$A$" (proportional to the maximal polarization of polydomain state). Constants $q_1$ and $q_2$ are given by Eq. (B.6c). These functions satisfy the boundary conditions (B.7a) and (B.7e) for arbitrary values of wave number $k$, while the condition (B.7c) is satisfied under the condition (B.11).

The amplitudes "$P$" and "$A$" are the variational parameters that could be determined from the minimization of the free energy functional with trial functions (B.16). Below we use much simpler trial function, since $q_1 \approx i\frac{\pi}{2h}$ (see Eq.(B.14f)) for a thick dielectric layer and $q_2 h \gg 1$ (see Eq.(B.6d))

$$P_3 \approx P + 2A \cos(kx) \left[ \cosh(q_1(z-h)) - \frac{q_1 \sinh(q_1 h)}{q_2} \exp(-q_2 z) \right], \quad \text{(B.17a)}$$

$$\phi^{(in)} \approx \frac{-P \frac{d}{\varepsilon_e}(h-z)}{\varepsilon_0 \varepsilon_b \frac{h}{\varepsilon_b}+\frac{d}{\varepsilon_e}} + \frac{2A}{\varepsilon_0 \varepsilon_b} \cos(kx) \left[ \frac{q_1}{q_1^2 - k^2} \sinh(q_1(z-h)) - \frac{q_1 \sinh(q_1 h)}{q_2^2 - k^2} \exp(-q_2 z) \right] \quad \text{(B.17b)}$$

In the most cases the terms proportional to $\exp(-q_2 z)$ give only the slightest corrections to the average values:

$$\left\langle \left[ 2\cos(k\,x) \left\{ \cos\left( \frac{\pi}{2h}(z-h) \right) + \frac{\pi}{2 q_2 h} \exp(-q_2 z) \right\} \right]^2 \right\rangle \approx 1 + \frac{5\pi^2}{4 h^3 q_2^3}, \quad \text{(B.18a)}$$

while their presence is important for the fulfillment of the boundary conditions. Keeping these arguments in mind, one could substitute Eqs. (B.17) to the functional (A.1) in order to obtain the following expression

$$\Delta G = \left\langle \frac{\alpha}{2} P_3^2 + \frac{\beta}{4} P_3^4 - E_{ext} P_3 + \frac{g_{11}}{2} \left( \frac{\partial P_3}{\partial z} \right)^2 + \frac{g_{44}}{2} \left( \frac{\partial P_3}{\partial x} \right)^2 + \frac{1}{2} P_3 \frac{\partial \phi}{\partial z} \right\rangle \approx$$

$$\approx \frac{\alpha_p}{2} P^2 + \frac{\alpha_a}{2} A^2 + \frac{\beta}{4} \left( P^4 + 6 P^2 A^2 + \frac{9}{4} A^4 \right) - E_{eff} P. \quad \text{(B.18b)}$$

Here we introduced the renormalized coefficients and effective electric field:

$$\alpha_p = \alpha + \frac{d}{\varepsilon_0(d\varepsilon_b + h\varepsilon_e)}, \tag{B.19a}$$

$$\alpha_a = \alpha - g_{11}q_1^2 + g_{44}k^2 - \frac{1}{\varepsilon_0\varepsilon_b}\frac{q_1^2}{k^2-q_1^2} \approx \alpha + g_{11}\left(\frac{\pi}{2h}\right)^2 + g_{44}k^2 + \frac{1}{\varepsilon_0\varepsilon_b}\frac{1}{1+(2kh/\pi)^2}, \tag{B.19b}$$

$$E_{eff} = \frac{\varepsilon_e}{d\varepsilon_b + h\varepsilon_e}V \tag{B.19c}$$

Equations of state are obtained from the minimization of (B.18b):

$$(\alpha_p + 3\beta A^2)P + \beta P^3 = E_{eff}, \tag{B.20a}$$

$$(\alpha_a + 3\beta P^2)A + \frac{9}{4}\beta A^3 = 0. \tag{B.20b}$$

The solution of the system (B.20) allows one to determine the possible phases as it shown below:

I. Paraelectric (PE) phase with

$$A = 0 \quad \text{and} \quad P \cong \frac{E_{eff}}{\alpha_p}, \tag{B.21a}$$

where $P = 0$ at $E_{eff} = 0$. It is stable at

$$\alpha_p > 0 \quad \text{and} \quad \alpha_a + 3\beta P^2 > 0. \tag{B.21b}$$

The energy of the PE phase is

$$\Delta G = 0 \tag{B.22}$$

II. Single domain ferroelectric (SDFE) phase with

$$A = 0 \quad \text{and} \quad P = \pm\sqrt{\frac{-\alpha_p}{\beta}} \quad \text{at } E_{eff} = 0. \tag{B.23}$$

is stable under the conditions

$$\alpha_p < 0 \text{ and } \alpha_a + 3\beta P^2 > 0 \tag{B.24a}$$

or

$$\alpha_p < 0 \text{ and } \alpha_a - 3\alpha_p > 0 \quad \text{at } E_{eff} = 0. \tag{B.24b}$$

The critical value of electric field could be obtained from the second condition (B.24a) and equation of state (B.20a):

$$E_{eff} > E_{cr}^{(f)} = \left(\alpha_p - \frac{\alpha_a}{3}\right)\sqrt{\frac{-\alpha_a}{3\beta}} \tag{B.25}$$

Energy of the SDFE phase is shown below

$$\Delta G = -\frac{\alpha_p}{2}P^2 - 3\frac{\beta}{4}P^4 = \left|\text{at } E_{eff} = 0\right| = -\frac{\alpha_p^2}{4\beta} \tag{B.26}$$

III. "pure" polydomain ferroelectric (PDFE) phase is possible only under at $E_{eff} = 0$ with

$$A = \pm\frac{2}{3}\sqrt{\frac{-\alpha_a}{\beta}} \text{ and } P = 0 \text{ (at } E_{eff} = 0). \tag{B.27a}$$

It is stable under the conditions that

$$\alpha_p + 3\beta A^2 = \alpha_p - \frac{4}{3}\alpha_a > 0 \text{ and } \alpha_a < 0 \text{ (at } E_{eff} = 0\text{)} \quad \text{(B.27b)}$$

Energy of PDFE phase at $E_{eff}$ is

$$\Delta G = -\frac{\alpha_a^2}{9\beta}. \quad \text{(B.28)}$$

IV. Partially "polarized" polydomain modulated ferroelectric (MFE) phase, with a modulation amplitude

$$A = \pm \frac{2}{3}\sqrt{\frac{-\alpha_a - 3\beta P^2}{\beta}} \quad \text{(B.29a)}$$

and polarization satisfying the equation

$$\left(\alpha_p - \frac{4}{3}\alpha_a\right)P - 3\beta P^3 = E_{eff} \quad \text{(B.29b)}$$

Note that due to the negative sign before the second term of (B.29b), it actually gives inverted "hysteresis" curve. The solution is stable when the matrix below is positively defined

$$\begin{pmatrix} \alpha_p + 3\beta A^2 + 3\beta P^2 & 6\beta AP \\ 6\beta PA & \alpha_a + 3\beta P^2 + \frac{27}{4}\beta A^2 \end{pmatrix}. \quad \text{(B.30a)}$$

That is possible under the conditions

$$\alpha_p - \frac{4}{3}\alpha_a - \beta P^2 > 0, \quad \text{(B.30b)}$$

$$-2(\alpha_a + 3\beta P^2) > 0, \quad \text{(B.30c)}$$

$$\left(\alpha_p + 3\beta A^2 + 3\beta P^2\right)\left(\alpha_a + 3\beta P^2 + \frac{27}{4}\beta A^2\right) - 36\beta^2 A^2 P^2 =$$

$$= -2(\alpha_a + 3\beta P^2)\left[\alpha_p - \frac{4}{3}\alpha_a - 9\beta P^2\right] > 0 \quad \text{(B.30d)}$$

These three conditions (B.30b-d) are equivalent to the pair of the following conditions:

$$-2(\alpha_a + 3\beta P^2) > 0 \text{ and } \alpha_p - \frac{4}{3}\alpha_a - 9\beta P^2 > 0 \quad \text{(B.30e)}$$

Energy of MFE phase is

$$\Delta G = -\frac{\alpha_a^2}{9\beta} + \frac{1}{2}\left(\alpha_p - \frac{4\alpha_a}{3}\right)P^2 - \frac{3\beta}{4}P^4 - E_{eff}P \quad \text{(B.31)}$$

The effective dielectric susceptibility in MFE phase is

$$\frac{\partial P}{\partial E_{eff}} = \frac{\alpha_a + 3\beta P^2 + \frac{27}{4}\beta A^2}{\left(\alpha_p + 3\beta A^2 + 3\beta P^2\right)\left(\alpha_a + 3\beta P^2 + \frac{27}{4}\beta A^2\right) - 36\beta^2 A^2 P^2} = \frac{1}{\alpha_p - \frac{4}{3}\alpha_a - 9\beta P^2} \quad \text{(B.32)}$$

According to the stability conditions (B.30e), $\partial P/\partial E_{eff} > 0$ even in the presence of the domain structure. On the other hand, a rather high electric field could make this phase unstable with negative permittivity.

Critical fields are obtained from the condition of zero of $A$ parameter, which is achieved at $P = \sqrt{-\alpha_a/3\beta}$, which in turn, according to the equation (B.29b), corresponds to the field value of

$$E_{cr}^{(SDFE)} = \left(\alpha_p - \frac{\alpha_a}{3}\right)\sqrt{\frac{-\alpha_a}{3\beta}} \quad (B.33a)$$

Also, one could rewrite the second inequality from (B.30e) as the upper limit for polarization $9\beta P^2 < \alpha_p - 4\alpha_a/3$, giving rise to the upper limit for the electric field

$$E_{eff} < E_{cr}^{(MFE)} = \frac{2}{3}\left(\alpha_p - \frac{4}{3}\alpha_a\right)\sqrt{\frac{\alpha_p - \frac{4}{3}\alpha_a}{9\beta}} \quad (B.33b)$$

Is it possible to have "true" mixed phase, SDFE-PDFE, which is a rippled single domain phase, at zero field, $E_{eff} = 0$? Formally, the system (B.29) has nonzero solution at $E_{eff} = 0$,

$$A = \pm\frac{2}{3}\sqrt{\frac{\alpha_a - 3\alpha_p}{3\beta}}, \quad P = \pm\sqrt{\frac{3\alpha_p - 4\alpha_a}{9\beta}} \quad (B.34)$$

which is real under the conditions

$$\alpha_a - 3\alpha_p > 0 \text{ and } 3\alpha_p - 4\alpha_a > 0. \quad (B.35)$$

However, the stability matrix (B.30a) calculated with the solution (B.34)

$$\begin{pmatrix} \frac{2}{9}(3\alpha_p - 4\alpha_a) & \frac{4\sqrt{\alpha_a - 3\alpha_p}\sqrt{-4\alpha_a + 3\alpha_p}}{3\sqrt{3}} \\ \frac{4\sqrt{\alpha_a - 3\alpha_p}\sqrt{-4\alpha_a + 3\alpha_p}}{3\sqrt{3}} & \frac{2}{3}(\alpha_a - 3\alpha_p) \end{pmatrix} \quad (B.36)$$

Has a determinant $\frac{4}{9}(\alpha_a - 3\alpha_p)(4\alpha_a - 3\alpha_p)$, which is obviously negative under the conditions (B.35), which means that the solution (B.34) is unstable.

## Appendix C. The general problem statement for FEM

The system of equations for the polarization $P_3$ and electric potential of the ferroelectric film $\phi^{(in)}$ and of the dielectric layer $\phi^{(out)}$ has the following form

$$\alpha P_3 + \beta P_3^3 + \gamma P_3^5 - g_{11}\frac{\partial^2 P_3}{\partial z^2} - g_{44}\left(\frac{\partial^2 P_3}{\partial x^2} + \frac{\partial^2 P_3}{\partial y^2}\right) = -\frac{\partial \phi^{(in)}}{\partial z} \quad \text{at } z \in [0, h] \quad (C.1a)$$

$$\left(\frac{\partial^2}{\partial x^2} + \frac{\partial^2}{\partial y^2} + \frac{\partial^2}{\partial z^2}\right)\phi^{(in)} = \frac{1}{\varepsilon_0 \varepsilon_b}\frac{\partial P_3}{\partial z}, \quad \text{at } z \in [0, h] \quad (C.1b)$$

$$\left(\frac{\partial^2}{\partial x^2} + \frac{\partial^2}{\partial y^2} + \frac{\partial^2}{\partial z^2}\right)\phi^{(out)} = 0. \quad \text{at } z \in [-d, 0] \quad (C.1c)$$

Let us consider the boundary and interface conditions at the different physical surfaces/interfaces:

$$\left.\left(\frac{\partial P_3}{\partial z}\right)\right|_{z=0,h} = 0, \quad (C.2a)$$

$$\left.\phi^{(out)}\right|_{z=-d} = V \quad (C.2b)$$

$$\left(-\varepsilon_0\varepsilon_b \frac{\partial \phi^{(in)}}{\partial z} + P_3 + \varepsilon_0\varepsilon_e \frac{\partial \phi^{(out)}}{\partial z}\right)\bigg|_{z=0} = 0. \quad \text{(C.2c)}$$

$$\left(\phi^{(out)} - \phi^{(in)}\right)\big|_{z=0} = 0 \quad \text{(C.2d)}$$

$$\phi^{(in)}\big|_{z=h} = 0 \quad \text{(C.2e)}$$

Here V is the bias voltage between the top and bottom electrodes. In order to simulate a laterally infinite system, we impose periodic boundary conditions at the virtual boundaries of computational box ($x = \pm H_x$ and $y = \pm H_y$):

$$P_3|_{x=-H_x} = P_3|_{x=H_x} \quad \text{(C.3a)}$$

$$\phi^{(in)}\big|_{x=-H_x} = \phi^{(in)}\big|_{x=H_x} \quad \text{(C.3b)}$$

$$\phi^{(out)}\big|_{x=-H_x} = \phi^{(out)}\big|_{x=H_x} \quad \text{(C.3c)}$$

and

$$P_3|_{y=-H_y} = P_3|_{y=H_y} \quad \text{(C.4a)}$$

$$\phi^{(in)}\big|_{y=-H_y} = \phi^{(in)}\big|_{y=H_y} \quad \text{(C.4b)}$$

$$\phi^{(out)}\big|_{y=-H_y} = \phi^{(out)}\big|_{y=H_y} \quad \text{(C.4c)}$$

### Appendix D. Bifurcation analysis of the nonlinear system (8)

Let us assume that the screening degree is high enough, so that $\left|\frac{e}{k_BT}\frac{hd(\sigma_p+\sigma_n-\bar{P})}{\varepsilon_0(\varepsilon_e h+d\varepsilon_b)}\right| \ll 1$. In the case we can expand the expressions for $\sigma_{p0}[\Psi]$ and $\sigma_{n0}[\Psi]$ in series allowing for Eq.(8b) and the following expansion:

$$\frac{eZ_i}{A_i}\left(1 + g_i \exp\left(\frac{\Delta G_i^0 + eZ_i\Psi}{k_BT}\right)\right)^{-1} \approx \frac{eZ_i}{A_i}\frac{1}{1+S_i(\rho,T,V)} - \frac{hd(\sigma_p+\sigma_n-\bar{P})}{\varepsilon_0(\varepsilon_e h+d\varepsilon_b)}Q_i(\rho,T,V) \quad \text{(D.1a)}$$

where the subscript $i = p, n$, and the positive functions are introduced:

$$S_i(\rho,T,V) = g_i\exp\left(\frac{\Delta G_i^0}{k_BT} + \frac{eZ_i}{k_BT}\frac{\varepsilon_e hV}{\varepsilon_e h+d\varepsilon_b}\right), \quad Q_i(\rho,T,V) = \frac{(eZ_i)^2}{A_i k_BT}\frac{S_i(\rho,T,V)}{(1+S_i(\rho,T,V))^2}. \quad \text{(D.1b)}$$

Below we omit the arguments $\rho, T, V$ in the functions $S_i(\rho,T,V)$ and $Q_i(\rho,T,V)$ for brevity. Consequently Eqs.(8a)-(8d) become the system of coupled differential equations for three variables $\bar{P}, \sigma_p$ and $\sigma_n$:

$$\Gamma\frac{d}{dt}\bar{P} + \left(\alpha_T(T-T_c) + \frac{d}{\varepsilon_0(\varepsilon_b d+\varepsilon_e h)}\right)\bar{P} + \beta\bar{P}^3 + \gamma\bar{P}^5 - \frac{d(\sigma_p+\sigma_n)}{\varepsilon_0(\varepsilon_e h+d\varepsilon_b)} = \frac{\varepsilon_e V}{\varepsilon_e h+d\varepsilon_b}, \quad \text{(D.2a)}$$

$$\tau_p\frac{\partial \sigma_p}{\partial t} + \left(1 + \frac{hdQ_p}{\varepsilon_0(\varepsilon_e h+d\varepsilon_b)}\right)\sigma_p + \frac{hdQ_p}{\varepsilon_0(\varepsilon_e h+d\varepsilon_b)}(\sigma_n-\bar{P}) = \frac{eZ_p}{A_p}\frac{1}{1+S_p}, \quad \text{(D.2b)}$$

$$\tau_n\frac{\partial \sigma_n}{\partial t} + \left(1 + \frac{hdQ_n}{\varepsilon_0(\varepsilon_e h+d\varepsilon_b)}\right)\sigma_n + \frac{hdQ_n}{\varepsilon_0(\varepsilon_e h+d\varepsilon_b)}(\sigma_p-\bar{P}) = \frac{eZ_n}{A_n}\frac{1}{1+S_n}, \quad \text{(D.2c)}$$

The stationary solutions of the system (D.2) obey the equations:

$$\left(\alpha_T(T-T_c) + \frac{d}{\varepsilon_0(\varepsilon_b d + \varepsilon_e h)}\right)\bar{P}_S + \beta\bar{P}_S^3 + \gamma\bar{P}_S^5 = \frac{\sigma_S d + \varepsilon_0\varepsilon_e V}{\varepsilon_0(\varepsilon_e h + d\varepsilon_b)}. \tag{D.3a}$$

$$\sigma_S = \left[\frac{hd(Q_p+Q_n)}{\varepsilon_0(\varepsilon_e h + d\varepsilon_b)}\bar{P}_S + \frac{eZ_p}{A_p}\frac{1}{1+S_p} + \frac{eZ_n}{A_n}\frac{1}{1+S_n}\right]\left[1 + \frac{hd(Q_p+Q_n)}{\varepsilon_0(\varepsilon_e h + d\varepsilon_b)}\right]^{-1}, \tag{D.3b}$$

where $\sigma_S = \sigma_p + \sigma_n$. Substitution of expression (D.3b) into Eq.(D.3a) leads to the equation for polarization with renormalized coefficients:

$$\alpha_R \bar{P}_S + \beta\bar{P}_S^3 + \gamma\bar{P}_S^5 = E_S + E_{eff}, \tag{D.4a}$$

$$\alpha_R(\rho, T, V, h) = \alpha_T(T-T_c) + \frac{d}{\varepsilon_0(\varepsilon_e h + d\varepsilon_b) + hd(Q_p+Q_n)}, \tag{D.4b}$$

$$E_S(\rho, T, V) = \frac{d}{\varepsilon_0(\varepsilon_e h + d\varepsilon_b) + hd(Q_p+Q_n)}\left[\frac{eZ_p}{A_p}\frac{1}{1+S_p} + \frac{eZ_n}{A_n}\frac{1}{1+S_n}\right], \tag{D.4c}$$

$$E_{eff} = \frac{\varepsilon_e V}{\varepsilon_e h + d\varepsilon_b}. \tag{D.4d}$$

Let us perform a bifurcation analysis of the system (D.2) in the vicinity of the stationary points (D.4). Assuming that $\bar{P} = \bar{P}_S + \delta P \exp(\lambda t)$ and $\sigma_{n,p} = \sigma_{ns,ps} + \delta\sigma_{n,p}\exp(\lambda t)$, we obtain the determinant for $\lambda$ determination:

$$\begin{vmatrix} \lambda\Gamma + \alpha + \frac{d}{\varepsilon_0(\varepsilon_b d + \varepsilon_e h)} + 3\beta\bar{P}_S^2 + 5\gamma\bar{P}_S^4 & -\frac{d}{\varepsilon_0(\varepsilon_e h + d\varepsilon_b)} & -\frac{d}{\varepsilon_0(\varepsilon_e h + d\varepsilon_b)} \\ -\frac{hdQ_p}{\varepsilon_0(\varepsilon_e h + d\varepsilon_b)} & \lambda\tau_p + 1 + \frac{hdQ_p}{\varepsilon_0(\varepsilon_e h + d\varepsilon_b)} & \frac{hdQ_p}{\varepsilon_0(\varepsilon_e h + d\varepsilon_b)} \\ -\frac{hdQ_n}{\varepsilon_0(\varepsilon_e h + d\varepsilon_b)} & \frac{hdQ_n}{\varepsilon_0(\varepsilon_e h + d\varepsilon_b)} & \lambda\tau_n + 1 + \frac{hdQ_n}{\varepsilon_0(\varepsilon_e h + d\varepsilon_b)} \end{vmatrix} = 0$$

(D.5a)

That is equivalent to the cubic equation

$$\left(\lambda\Gamma + \alpha + \frac{d}{\varepsilon_0(\varepsilon_b d + \varepsilon_e h)} + 3\beta\bar{P}_S^2 + 5\gamma\bar{P}_S^4\right)(\lambda\tau_p + 1)(\lambda\tau_n + 1) + \frac{hd(\lambda\Gamma + \alpha + 3\beta\bar{P}_S^2 + 5\gamma\bar{P}_S^4)}{\varepsilon_0(\varepsilon_e h + d\varepsilon_b)}\left[(\lambda\tau_p + 1)Q_n + (\lambda\tau_n + 1)Q_p\right] = 0 \tag{D.5b}$$

Here we are looking for the instability threshold corresponding to condition $Re[\lambda] = 0$. Typically, the relaxation times of electrons ions is much smaller than for the ions, e.g., $\tau_n \ll \tau_p$. Far from the Curie temperature $\Gamma/|\alpha| \ll \tau_n \ll \tau_p$ and we can put $\Gamma = 0$ in Eq.(D.5b), making it a quadratic equation. Numerical estimates made for $A_n = A_p = A = 10^{-18}$ m$^2$, $Z_p = -Z_n = +1$, and $g_p = g_n = 1$ show that the condition $Re[\lambda] \geq 0$ can take place in many cases explaining the peak-type features in **Figs. 8-9** and turning points located near the coercive and zero voltages in **Fig. 10** in the main text.